%% file: sigconf.tex
\documentclass[sigconf]{acmart}
\usepackage{multirow}
\usepackage{graphicx, caption, subcaption}
\usepackage{enumitem}
\usepackage[normalem]{ulem}
\useunder{\uline}{\ul}{}
\usepackage[ruled]{algorithm2e}

\SetCommentSty{mycommfont}
\usepackage{array}
\usepackage{booktabs}
\usepackage{bbding}
\setlist[itemize]{leftmargin=*}

\AtBeginDocument{%
  \providecommand\BibTeX{{%
    \normalfont B\kern-0.5em{\scshape i\kern-0.25em b}\kern-0.8em\TeX}}}

\copyrightyear{2022}
\acmYear{2022}
\setcopyright{rightsretained}
\acmConference[CIKM '22]{Proceedings of the 31st ACM International Conference on Information and Knowledge Management}{October 17--21, 2022}{Atlanta, GA, USA}
\acmBooktitle{Proceedings of the 31st ACM Int'l Conference on Information and Knowledge Management (CIKM '22), Oct. 17--21, 2022, Atlanta, GA, USA}
\acmISBN{978-1-4503-9236-5/22/10}
\acmDOI{10.1145/3511808.3557289}

\usepackage{etoolbox}
\makeatletter
\patchcmd{\maketitle}{\@copyrightpermission}{
  \begin{minipage}{0.3\columnwidth}
     \href{https://creativecommons.org/licenses/by/4.0/}{\includegraphics[width=0.90\textwidth]{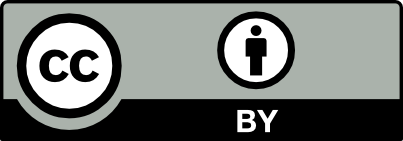}}
  \end{minipage}\hfill
  \begin{minipage}{0.7\columnwidth}
     \href{https://creativecommons.org/licenses/by/4.0/}{This work is licensed under a Creative Commons Attribution International 4.0 License.}
  \end{minipage}
  
  \vspace{5pt}
}{}{}

\makeatother

\begin{document}

\title{Disentangling Past-Future Modeling in Sequential Recommendation via Dual Networks}

\author{Hengyu Zhang}
\email{zhang-hy21@mails.tsinghua.edu.cn}
\authornote{Work done when they were research interns at Huawei Noah’s Ark Lab, and both authors contributed equally to this research.}
\affiliation{
  \institution{Tsinghua Shenzhen International Graduate School, Tsinghua University}
  \city{Shenzhen}
  \country{China}}

\author{Enming Yuan}
\authornotemark[1]
\email{yem19@mails.tsinghua.edu.cn}
\affiliation{
  \institution{Institute for Interdisciplinary Information Sciences,Tsinghua University}
  \city{Beijing}
  \country{China}}

\author{Wei Guo}
\email{guowei67@huawei.com}
\affiliation{
  \institution{Huawei Noah's Ark Lab}
  \city{Shenzhen}
  \country{China}}

\author{Zhicheng He}
\email{hezhicheng9@huawei.com}
\affiliation{
  \institution{Huawei Noah's Ark Lab}
  \city{Shenzhen}
  \country{China}}

\author{Jiarui Qin}
\email{qinjr@icloud.com}
\affiliation{
  \institution{Shanghai Jiao Tong University}
  \city{Shanghai}
  \country{China}}

\author{Huifeng Guo}
\email{huifeng.guo@huawei.com}
\affiliation{
  \institution{Huawei Noah's Ark Lab}
  \city{Shenzhen}
  \country{China}}

\author{Bo Chen}
\email{chenbo116@huawei.com}
\affiliation{
  \institution{Huawei Noah's Ark Lab}
  \city{Shenzhen}
  \country{China}
 }

\author{Xiu Li}
\email{li.xiu@sz.tsinghua.edu.cn}
\authornote{Corresponding author.}
\affiliation{
  \institution{Tsinghua Shenzhen International Graduate School, Tsinghua University}
  \city{Shenzhen}
  \country{China}}

\author{Ruiming Tang}
\email{tangruiming@huawei.com}
\authornotemark[2]
\affiliation{
  \institution{Huawei Noah's Ark Lab}
  \city{Shenzhen}
  \country{China}}

\renewcommand{\shortauthors}{Hengyu Zhang, et al.}

\begin{abstract}
Sequential recommendation (SR) plays an important role in personalized recommender systems because it captures dynamic and diverse preferences from users' real-time increasing behaviors.
Unlike the standard autoregressive training strategy, future data (also available during training) has been used to facilitate model training as it provides richer signals about user's current interests and can be used to improve the recommendation quality.
However, these methods suffer from a severe training-inference gap, i.e., both past and future contexts are modeled by the same encoder when training, while only historical behaviors are available during inference. 
This discrepancy leads to potential performance degradation.
To alleviate the training-inference gap, we propose a new framework DualRec, which achieves past-future disentanglement and past-future mutual enhancement by a novel dual network.
Specifically, a dual network structure is exploited to model the past and future context separately. And a bi-directional knowledge transferring mechanism enhances the knowledge learnt by the dual network.
Extensive experiments on four real-world datasets demonstrate the superiority of our approach over baseline methods. 
Besides, we demonstrate the compatibility of DualRec by instantiating using RNN, Transformer, and filter-MLP as backbones.
Further empirical analysis verifies the high utility of modeling future contexts under our DualRec framework.
The code of DualRec is publicly available at https://github.com/zhy99426/DualRec.

\end{abstract}



\maketitle

\input{introduction}

\input{related_work}

\input{method}

\input{experiments}

\section{CONCLUSIONS}

In this paper, we present the DualRec framework to introduce future data into the training process while alleviating the potential training-inference gap. 
Specifically, a dual network is proposed to achieve past-future disentanglement, hence avoiding training-inference gap. 
Further, bi-directional information transferring is devised to make better use of both past and future context knowledge.
Extensive experiments on four real-world datasets demonstrate the superior effectiveness and wide compatibility of the DualRec framework. 
In future work, we will further explore adopting our DualRec framework to the multi-behavior sequential recommendation.

\begin{acks}

This work was partly supported by the Science and Technology Innovation 2030-Key Project under Grant 2021ZD0201404 and Aminer·
ShenZhen·ScientificSuperBrain. 
And we thank MindSpore\cite{mindspore} for the partial support of this work, which is a new deep learning computing framework.

\end{acks}

\bibliographystyle{ACM-Reference-Format}
\balance
\bibliography{ref}

\end{document}

%% file: introduction.tex
\section{Introduction}
\label{sec:intro}

Recommender systems have been widely deployed in online service platforms, ranging from online advertising and retailing \cite{wide-deep,deepfm,wang2018billion} to music and video recommendation \cite{van2013music,davidson2010youtube,deldjoo2016video}. 
Generally, users' interests are dynamic and evolve over time, which are depicted by users' sequential interactions. 
Therefore, it leads to Sequential Recommendation (SR) that models the sequential characteristics of users' behaviors and provides more precise and customized services.
To make accurate predictions, it's essential to learn effective representation for users based on the historical interactions they have engaged with.
Over the years, great efforts have been devoted, and different model architectures are proposed for sequential recommendation, including Recurrent Neural Network (RNN) \cite{hidasi2015session}, Convolutional Neural Networks (CNN) \cite{tang2018personalized}, Self-Attention Network (SAN)\cite{kang2018self}, and Graph Neural Networks (GNN) \cite{qiu2019rethinking,wu2019session}.

\begin{figure}[t]
    \centering
    \includegraphics[width=0.47\textwidth]{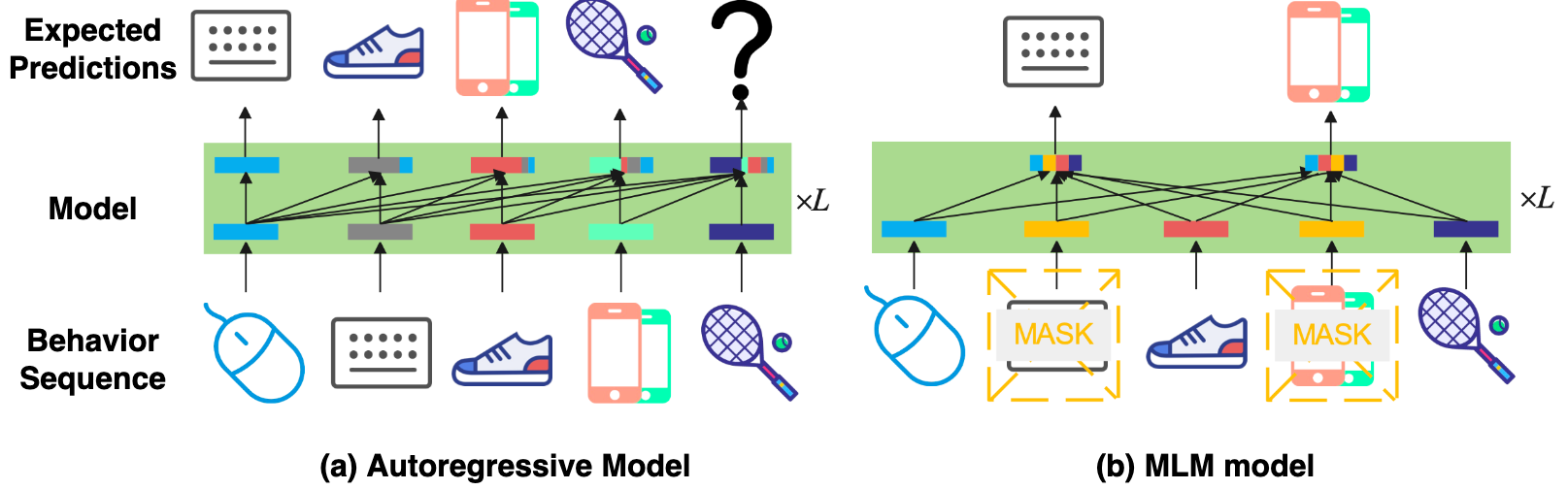}
    \caption{Illustration of common sequential recommendation models.}
    \label{fig:illustration}
    \vspace{-5mm}
\end{figure}

Typically, the sequential recommendation problem is formulated as a \textit{next-item-prediction} problem, also known as autoregressive model (Figure \ref{fig:illustration}a), which predicts the next item a user will interact with based on her historical behaviors \cite{tang2018personalized,hidasi2015session,kang2018self}.
It is a straightforward modeling choice for sequential data.
However, in the sequential recommendation, the autoregressive schema could weaken the model's expressiveness.
Because in practice, the sequential dependencies of user behaviors may not strictly hold.
For example, after purchasing an iPad, a user may click on Apple pencil, iPad case, and headphones. 
But it is likely that the user clicked on these three products at random.
Then simply modeling it in a compulsory sequential order loses some overall contextual information.
Because future data (interactions that occur after the target interaction) also provide rich collaborative information to assist the model training.
Therefore, it's reasonable to leverage the future data to train better sequential recommendation models.

Recently, researchers have proven that leveraging both past and future contextual information during training will significantly boost recommendation performances compared to the autoregressive models \cite{sun2019bert4rec, yuan2020future}.
For example, inspired by the advances in the field of natural language processing (NLP), BERT4Rec \cite{sun2019bert4rec} employs a masked language model (MLM) training objective which predicts masked items based on both historical and future behavior records during training (Figure \ref{fig:illustration}b).
BERT4Rec significantly improves recommendation performances compared to its unidirectional autoregressive counterpart SASRec \cite{kang2018self}.

Despite the richer contextual information brought by training with future interaction data, simply adopting MLM objectives for SR can introduce a severe \textbf{training-inference gap}.
Specifically, at training time, the MLM model predicts masked items with both past and future interactions as context, which can be illustrated as $P(i|\mathbf{x}_{past},\mathbf{x}_{future})$.
While at inference, only past behaviors are available available for prediction, i.e., $P(i|\mathbf{x}_{past},\texttt{NULL})$.
The discrepancy of context between training and inference can bias the model during inference and lead to potential performance degradation.

To exploit richer contextual information from the future while alleviating the potential training-inference gap. The following model desiderata should be met:

\begin{itemize}
    \item \textbf{Past-future disentanglement}:
    The training-inference gap in existing methods is caused by the use of a single encoder predictor that entangles past and future contextual information, thus messing with inference.
    Instead, the future data should be modeled in a separate way without explicitly interfering with modeling historical interaction data.
    If both disentangled encoders get well-trained, the absence of future information will not degrade the performance of the past information encoder.
    By this means, we can use only past behaviors for inference, with a minimal gap between training and inference.
    \item \textbf{Past-future mutual enhancement}:
    Users' interests captured by past and future behaviors are closely related and complementary.
    Simply separating past and future modeling processes hinders leveraging knowledge learned from each other.
    To better exploit future data, an elegant way is to have the disentangled past-future modeling process mutually enhance each other.
\end{itemize}

In this article, we propose a framework for better utilization of past and future information in sequential recommendation, named \textbf{DualRec}.
To alleviate the training-inference gap, DualRec adopts a dual network structure.
For a target interaction, past and future contextual behaviors are modeled by two encoders, respectively.
The two encoders perform dual tasks, i.e. the past encoder performs next-item prediction (primal) while the future encoder performs previous-item-prediction (dual).
Future information is decoupled from the modeling of past information in this way.
During inference, only the past encoder is used to make predictions, thus avoiding the training-inference gap.
Secondly, dual network enhance each other through a multi-scale knowledge transferring.
Specifically, the internal representations of two networks are constraint to alignment based on the assumption that users' interests captured by past and future behaviors are closely related and complementary.
Finally, as a general framework, DualRec can be instantiated using different backbone models, including RNN, Transformer and filter-based MLP.

To summarize, our contributions are as follows:
\begin{itemize}
    \item We highlight the training-inference gap existed in sequential recommendation models when leveraging future data.
    To handle this problem, we propose a novel framework DualRec that achieves the disentanglement and mutual enhancement of past-future modeling.
    \item DualRec explicitly decouples the past information and future information modeling into two separate encoders, thus alleviating the training-inference gap, and further using a past-future knowledge transferring to learn an enhanced representation.
    \item We conduct comprehensive experiments on four public datasets. Experimental results demonstrate the effectiveness of our proposed DualRec as compared with several baseline models. Further analysis illustrates its compatibility.
\end{itemize}

%% file: related_work.tex
\section{Related Work}
Sequential recommenders are designed to model the sequential dynamics in user behaviors.
Early efforts leverage the Markov Chain (MC) assumption \cite{2010Factorizing} and model item-item transition to predict user's next action based on the last visited items.
Recently, different neural network-based models have been applied, including Recurrent Neural Networks (RNNs) \cite{mikolov2010recurrent}, Convolutional Neural Networks (CNNs) \cite{krizhevsky2012imagenet}, Attention Networks \cite{vaswani2017attention} and Graph Neural Networks \cite{kipf2016gcn}. 
GRU4Rec \cite{hidasi2015session} is a pioneering work that employs RNN to capture the dynamic characteristic of user behaviors. 
\citeauthor{hidasi2018recurrent} further extends GRU4Rec with enhanced ranking functions as well as effective sampling strategies.
Another line of research is based on CNN.
Caser \cite{tang2018personalized} treats the embedding matrix of items as a 2D image and models user behavior sequences with convolution. 
The main advantage of CNN-based models is that they are much easier to be parallelized on GPUs compared with RNN-based models.
Self-Attention Mechanism and the Transformer architectures \cite{vaswani2017attention} are applied to sequential recommendation and proved to be advantageous in discovering user behavior patterns due to the adaptive weight learning and better integration of long-range dependencies, such as SASRec \cite{kang2018self}, BERT4Rec \cite{sun2019bert4rec} and S3-Rec \cite{zhou2020s3}.
More recently, Graph Neural Networks have been explored to encode the global structure of user interactions and capture complex item transitions. 
SRGNN \cite{wu2019session} and GCSAN \cite{xu2019graph} model sequence or session as graph structure, and use graph neural networks and attention mechanism to capture the rich dependencies. 
Inspired by work related to self-supervised learning, S3-Rec \cite{zhou2020s3} and CL4Rec \cite{xie2020contrastive} used a contrastive learning approach to pre-train the sequential recommendation model.
Furthermore, CLEA \cite{qin2021world} uses a contrastive learning approach to automatically filter out items from the user sequence that are more irrelevant to the target.

Among these sequential recommendation works, some bidirectional Transformer-based methods adopting MLM objectives, such as BERT4Rec\cite{sun2019bert4rec}, attempt to utilize both the past and future contextual information. However, these works suffer from a severe training-inference gap as analyzed in Section 
\ref{sec:intro}.

%% file: method.tex
\section{Method}\label{sec:method}
\begin{figure*}[!htbp]
    \centering
    \setlength{\abovecaptionskip}{0.1cm}
	\setlength{\belowcaptionskip}{0mm}
    \includegraphics[width=0.95\linewidth]{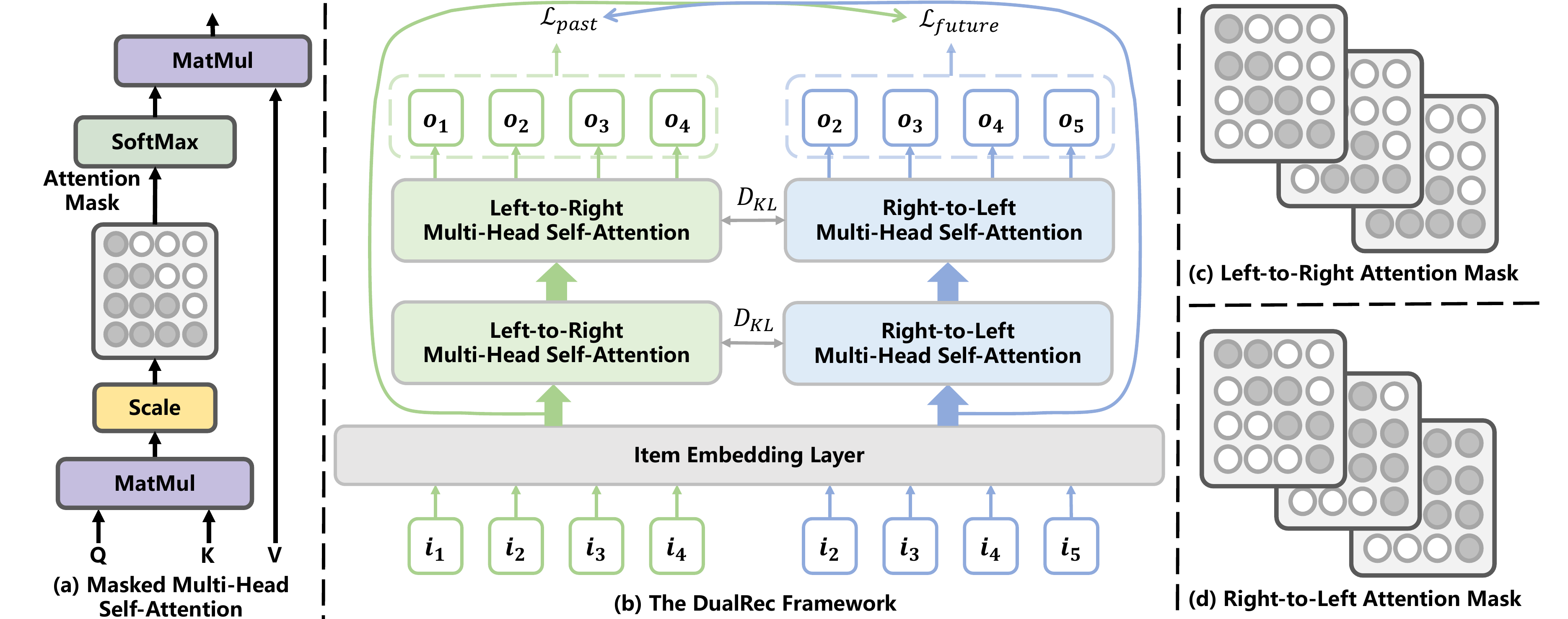}
    \caption{Model architecture of the proposed DualRec framework. (a) Illustration of masked multi-head self-attention. The shaded nodes are visible. (b) Overall structure of the dual network, with the past encoder on the left and the future encoder on the right. Past and future encoders are associated by the shared embedding layer and bi-directional information transferring using KL divergences. (c) and (d) are the attention masks from left to right in the past encoder and from right to left in the future encoder, respectively, and the time windows corresponding to each attention head are different, i.e., multi-scale. }
    \label{fig:dualrec}
\end{figure*}

In this section, we first define the sequential recommendation problem (Section \ref{sec:problem_formulation}). Then, we elaborate on the technical details of our proposed DualRec framework.
The DualRec framework is shown in Figure \ref{fig:dualrec} (b).
We first introduce the base encoder (Section \ref{sec:basemodel}), which is a Transformer-based backbone.
Then to utilize future context while alleviating the training-inference gap, we present the dual-network structure (Section \ref{sec:dual-network}) that models past and future behaviors through two encoders, respectively.
Furthermore, the bi-directional information transferring mechanism (Section \ref{sec:bi-direct}) is adopted to make the dual networks enhance each other.
The details in model training and inference are presented in Section \ref{sec:model_training}.
Further, we discuss the complexity and compatibility of DualRec (Section \ref{sec:model_discussion}).
Essentially, DualRec can work in cooperation with most existing sequential recommendation models.

\subsection{Problem Formulation}
\label{sec:problem_formulation}
Sequential recommendation learns to predict users' next behavior from their historical behavior sequences. 
Given a set of users $\mathcal{U}=\{u_1, u_2, ... , u_{|\mathcal{U}|}\}$, and a set of items $\mathcal{I}=\{i_1, i_2, ... , i_{|\mathcal{I}|}\}$, let $\mathcal{S}^{(u)}=[i_1^{(u)}, i_2^{(u)}, ... , i_{T_u}^{(u)}]$ denote the chronologically sorted behavior sequence of user $u \in \mathcal{U}$, where $i_t^{(u)}$ is $t$-th interacted item of user $u$, and $T_u=|\mathcal{S}^{(u)}|$ is the length of behavior sequence. 
A sequential recommendation model predicts the next item $i_{T_u+1}^{(u)}$ that user $u$ will interact with based on the behavior history $\mathcal{S}^{(u)}$, which can be formulated as:
\begin{equation}
p(i_{T_u+1}^{(u)} = i^{(c)}|\mathcal{S}^{(u)}) = \operatorname {SeqRecModel}(\mathcal{S}^{(u)}, i^{(c)}),
\end{equation}
where $i^{(c)}$ is the candidate item, and $\operatorname{SeqRecModel}(\cdot, \cdot)$ is a sequential recommendation model. 

\subsection{Base Encoder}
\label{sec:basemodel}
For simplicity, we illustrate with the standard Transformer as the base encoder of the dual network structure, which is widely used in many sequential recommendation methods \cite{kang2018self,sun2019bert4rec,zhou2020s3} and has been proven to be advantageous in discovering sequential patterns due to the adaptive weight learning.
Notably, our proposed framework can also work with other backbone architectures, including RNNs and CNNs.
We will evaluate the performance of our framework with different backbones in the experimental section.

\subsubsection{Embedding Layer.}
The input user behavior sequence $S^{(u)}$ is first transformed into a fixed-length sequence $s=(i_1, i_2, ... , i_n)$ (For simplicity, we omit the superscript $u$),
where $n$ is the predefined maximum length.
Specifically, if the original sequence length is greater than $n$, we keep the most recent $n$ actions. 
If the original sequence length is less than $n$, special \texttt{[padding]} tokens are padded to the left of the sequence as dummy past interactions.

\textbf{Item Embedding:}
For all items given in $\mathcal{I}$, we create an item embedding matrix $\mathbf{E}^{\mathcal{I}} \in \mathbb{R}^{\mid \mathcal{I} \mid \times d}$ where $d$ is the embedding size. 
User behavior sequence $s=(i_1, i_2, ... , i_n)$ is embedded as:
\begin{equation}\label{eq:item_embed}
\mathbf{X}^{(0)} = (\mathbf{e}_{1}, \cdots, \mathbf{e}_{n}), \ \mathbf{e}_{k} = \operatorname{LookUp}(i_{k}, \mathbf{E}^{\mathcal{I}}),
\end{equation}
where $\operatorname{LookUp}(\cdot, \cdot)$ retrieves an item embedding from the embedding matrix, and always gets a constant zero vector for the padding item.

\textbf{Relative Positional Embedding:}
Transformer doesn't contain any recurrent or convolutional operation and is not sensitive to positional information.
Therefore, we need to inject additional positional embedding.
We adopt the relative positional embedding as it is more robust than absolute positional embedding \cite{Shaw2018Self}.
When calculating the attention weight of the $j$-th item given the $i$-th item as query, the relative positional embedding is calculated as
\begin{equation}\label{eq:pos_embed}
    \mathbf{p}(i, j) = \operatorname{LookUp}(\operatorname{Dist}(i, j), \mathbf{E}^{\mathcal{P}}),
\end{equation}
where $\operatorname{Dist}(i, j) \in [-n+1, n-1]$ is the relative distance between two items.
$\mathbf{E}^{\mathcal{P}}\in \mathbb{R}^{(2n-1) \times h}$ is the positional embedding where $h$ is the number of heads in the self-attention mechanism.
The relative positional matrix $\mathbf{p}$ is added to the attention matrix to inject additional relative positional information.
 
\subsubsection{Transformer Layer}

After the embedding layer, the input $\mathbf{X} \in \mathbb{R}^{n \times d}$ are fed to multi-head self-attention (MSA) blocks for capturing the relations among different items in this sequence.
The computation for each head can be formulated as follows:
\begin{equation}
    \mathbf{head}_{i} = softmax\left ( \frac{(\mathbf{X}\mathbf{W}^Q_{i}) \cdot (\mathbf{X}\mathbf{W}_{i}^K)^{\top}}{\sqrt{d/h}} \right ) (\mathbf{X}\mathbf{W}^V_{i}),
\end{equation}
where $\mathbf{W}^Q_{i} \in \mathbb{R}^{d \times \frac{d}{h}}$, $\mathbf{W}^K_{i} \in \mathbb{R}^{d \times \frac{d}{h}}$ and $\mathbf{W}^V_{i} \in \mathbb{R}^{d \times \frac{d}{h}}$ are the query, key, and value projection matrices, respectively, $(\cdot)$ is the matrix multiplication operator, $i$ indicates a specific head, $h$ is the total number of heads and $\sqrt{d/h}$ is the scaling factor. 

MSA performs the above  self-attention operation $h$ times in parallel, then combines the outputs of each head together, and linearly projects them to get the final output: 
\begin{equation}
    \text{MSA}(\mathbf{X}) = \text{Concat}(\mathbf{head}_{i}, \mathbf{head}_{2}, ..., \mathbf{head}_{h}) \mathbf{W}^o,
\end{equation}
where $\mathbf{W}^o \in \mathbb{R}^{d \times d}$ is the corresponding transformation matrix.

To introduce non-linearity and perform feature transformation between MSA layers, point-wise feed-forward (PFF) layer is used:
\begin{equation}
    \text{PFF}(\mathbf{X}) = \text{FC}(\sigma(\text{FC}(\mathbf{X}))), \text{FC}(\mathbf{X}) = \mathbf{XW}+\mathbf{b},
\end{equation}
where $\mathbf{W} \in \mathbb{R}^{d \times d}$ and $\mathbf{b} \in \mathbb{R}^{1 \times d}$ are the weights and bias of a fully-connected (FC) layer, and $\sigma(\cdot)$ is the non-linear activation function such as ReLU \cite{glorot2011deep}.

Finally, residual connections \cite{he2016deep} and layer normalizations \cite{ba2016layer} are employed to connect the MSA and PFF modules to get the Transformer layer:

\begin{equation}
\small
\begin{aligned}
    \mathbf{H}^{(l)} &= \text{LayerNorm}\left(\mathbf{X}^{(l-1)}+\text{MSA}(\mathbf{X}^{(l-1)})\right), \\
    \mathbf{X}^{(l)} &= \text{LayerNorm}\left(\mathbf{H}^{(l)}+\text{PFF}(\mathbf{H}^{(l)})\right),
\end{aligned}
\end{equation} 
where $\mathbf{H}^{(l)}$ is the intermediate representation at layer $l$ and $\mathbf{X}^{(l)}$ is the hidden representation at layer $l$.

\subsection{Dual Network}\label{sec:dual-network}
To exploit future information while alleviating the potential training-inference gap, we present the dual network as shown in Figure \ref{fig:dualrec}(b).
In the dual network model, two encoders (denoted as past encoder and future encoder, respectively) are instantiated using the base encoder, through which the past and future modeling are disentangled explicitly.
For a target item, the two encoders perform dual tasks, i.e., the past encoder performs next-item-prediction (primal) while the future encoder performs previous-item-prediction (dual).
In this way, the future data is modeled in a separate way without explicitly interfering with the modeling of historical interaction data.
Hence no significant training-inference gap is introduced.

Next, we elaborate on the details of how we implement the dual network model, including a shared embedding layer connecting the two encoders and the dual self-attention mechanism that achieves the dual tasks.

\subsubsection{Shared Embedding Layer}
Embeddings play a vital role in sequential recommendation and determine the effectiveness of modeling associations between items.
Here, to improve the effectiveness of embedding learning, we share the parameters of the embedding layer in the past and future encoders.
Hence, the embeddings are learned by receiving knowledge from both past and future.

\subsubsection{Dual Self-Attention}
 
In the dual task setting, each encoder is a unidirectional autoregressive model, where the past encoder learns the past-to-future autoregressive function while the future encoder learns the opposite future-to-past autoregressive function. 
To this end, we employ the causal attention mask to ensure causality \cite{kang2018self, vaswani2017attention}.
As illustrate in Figure \ref{fig:dualrec} (a), we modify the attention by disabling attention links between $\mathbf{Q}_i$ and $\mathbf{K}_j (j>i)$ for the past encoder.
Likewise, we disable attention links between $\mathbf{Q}_i$ and $\mathbf{K}_j (j<i)$ for the future encoder.

\subsection{Bi-directional Information Transferring}\label{sec:bi-direct}
The dual network presented above computes representations for items with past and future information independently, which overlooks the interaction between the two encoders. 
However, users' preferences captured by past and future encoders are closely related and complementary.
Therefore, the modeling of past and future behaviors can be mutually enhanced.
Furthermore, in practice, user interests are multi-scale, as there are often both stable long-term interests and dynamic short-term interests.
Therefore, how to effectively model user interest at different scales and transfer useful knowledge to assist the learning of each encoder remains a challenging problem.
In this section, two encoders in the dual network are further endowed with the ability to extract users' multi-scale interests and transfer related and complementary knowledge to enhance each other.

\subsubsection{Multi-scale Interest Extraction}
User's interests can be divided into local interests, global interests and interests at other scales \cite{dien}.
Existing methods only consider short- and long-term interests in the past behaviors \cite{hu2020graph, zhu2020modeling}.
Here, users' multi-scale interests from both historical and future behaviors are considered.
To this end, we propose a multi-scale scheme that extracts users' interest representations at different temporal scales.
The multi-scale scheme can cooperate with the multi-head self-attention mechanism by adding multi-scale masks at different heads, thereby aggregating information with different receptive fields.
Specifically, we design heuristic rules to assign a window size to each head based on the head id $i$ and sequence length $n$:
\begin{equation}
WS(i)=\begin{cases}
i+1 & \text { if } i \leq \frac{h}{2} \\
\frac{h}{2}+\left\lceil\frac{\exp (i-\frac{h}{2})}{\exp \frac{h}{2}} \cdot\left(n-\frac{h}{2}\right)\right\rceil & \text { if } i > \frac{h}{2}
\end{cases}, \quad i=1\cdots h
\end{equation}
where $WS(i)$ is the window size assign function for head $i$.
To comprehensively capture users' interests at multi-scale level, we set $h/2$ heads with linearly increasing local windows and $h/2$ heads with exponentially increasing global windows, respectively.

For a specific head with a window size of $\sigma$, we rewrite the multi-head self-attention mechanism to adapt to the multi-scale scheme: 

\begin{equation}
\small
\begin{aligned}
    head_{i,j}(\mathbf{X}, \sigma) &= softmax\left ( \frac{(\mathbf{X}\mathbf{W}^Q_{i})_j \cdot S_{j}\left(\mathbf{X}\mathbf{W}_i^K, \sigma \right)^{\top}}{\sqrt{d/h}} \right )\cdot S_{j}\left(\mathbf{X}\mathbf{W}_i^V, \sigma \right),\\
    S_{j}(\mathbf{X}, \sigma) &= \left[\mathbf{X}_{j-\sigma}, \cdots, \mathbf{X}_{j}\right],
    \end{aligned}
\end{equation}

 where $head_{i,j}$ indicates the $i$-th attention head vector in $j$-th position, and $\mathbf{X}_j$ means the $j$-th vector in the matrix $\mathbf{X}$.
The function $S_{j}(\mathbf{X}, \sigma)$ extracts the context of $\mathbf{X}_j$ in the range $\sigma$, where $\sigma>0$ in past encoders and $\sigma<0$ in future encoders.
In Figure \ref{fig:bit}, an illustration of the multi-scale self-attention mechanism is shown for $\sigma = 2,3,4$.

In the subsequent compatibility analysis, for the backbone of the non-transformer architecture, we used multi-scale input to implement multi-scale interest extraction.

\subsubsection{Interest-level Knowledge Transferring}
After extracting users' multi-scale interests from both past and future behaviors, we perform past-future knowledge transferring to make the two encoders enhance each other.
We borrow the idea of knowledge distillation \cite{hinton2015distilling,sanh2019distilbert} and introduce a regularization term that encourages the representation distributions of two encoders to be consistent with each other.
In this way, the related and complementary user interests captured by the two encoders are enhanced.

\begin{figure}[!htbp]
    \centering
    \setlength{\abovecaptionskip}{0mm}
	\setlength{\belowcaptionskip}{-4mm}
    \includegraphics[width=0.45\textwidth]{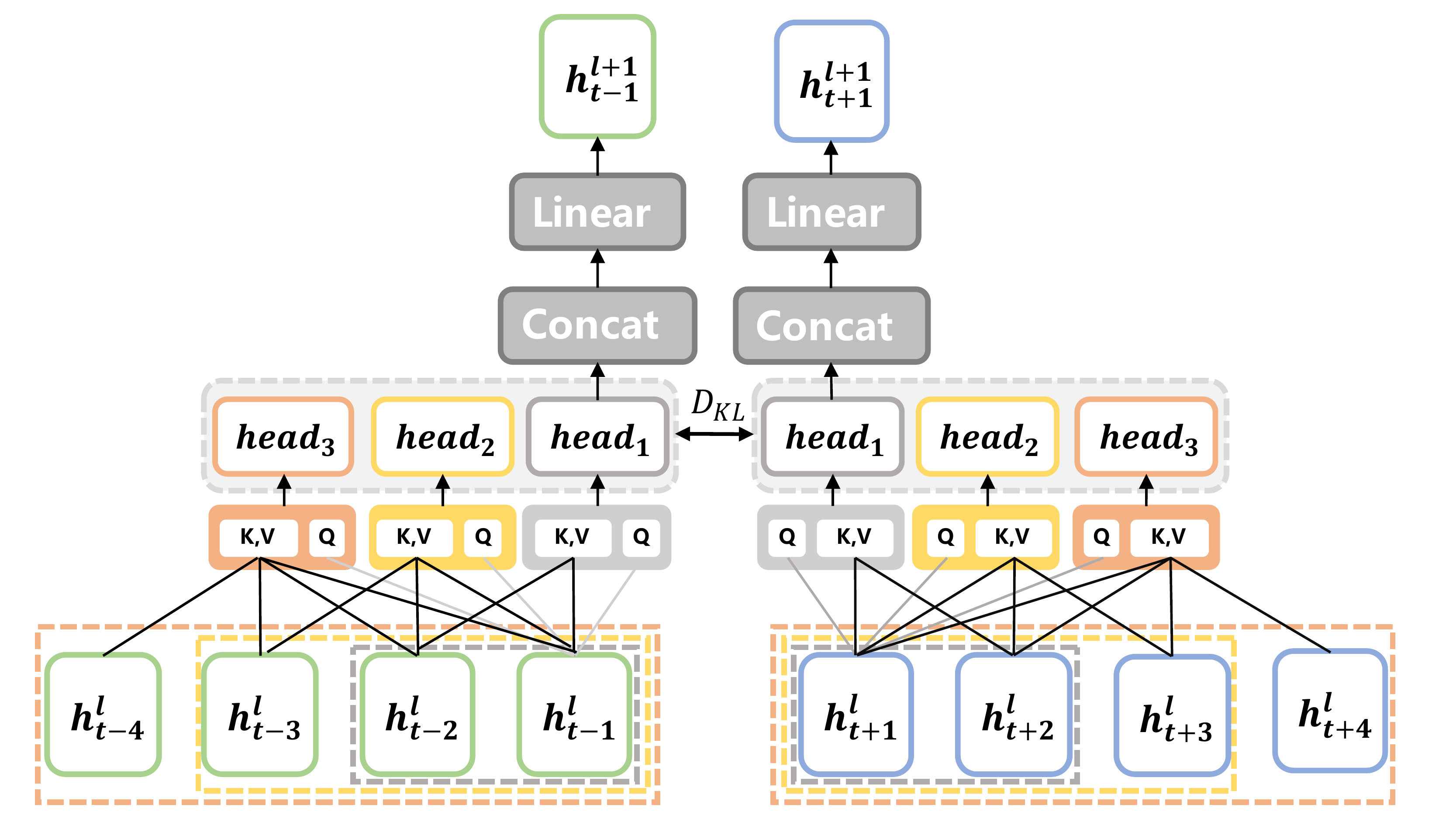}
    \caption{Illustration of interest-level knowledge transferring between past and future information.}
    \label{fig:bit}
\end{figure}

The goal is to regularize the model by minimizing the bidirectional Kullback-Leibler (KL) divergence between the past and future encoders' output distributions at multi-scale (Figure \ref{fig:bit}).
The regularization loss is calculated as:

\begin{equation}
    \mathcal{L}_{reg} = \sum_{i=1}^{h}\frac{1}{2}\left(D_{KL}(\textbf{head}^{p}_{i}||\textbf{head}^{f}_{i})+D_{KL}(\textbf{head}^{f}_{i}||\textbf{head}^{p}_{i})\right),
\end{equation} 
where $\textbf{head}^{p}_{i}$ and $\textbf{head}^{f}_{i}$ are output distributions at head $i$ of the past encoder and the future encoder, respectively.
And $D_{K L}(p \| q)$ is the KL divergence between two distributions:
\begin{equation}
D_{K L}(p \| q)=\sum_{i} p_i \log \frac{p_i}{q_i},
\end{equation}

\subsection{Model Training and Inference}
\label{sec:model_training}
\subsubsection{Model Training}
The two encoders in the dual network extract information of previously consumed items and future consumed items, respectively.
Correspondingly, when make prediction for the $t$-th behavior, we use the output $\mathbf{o}^{p}_{t-1}$ of the past encoder and $\mathbf{o}^{f}_{t+1}$ of the future encoder.
Specifically, we estimate the probability of each candidate item according to the inner product based similarity:

\begin{equation}
    \textbf{s}^{p}_{t} = \mathbf{o}^{p}_{t-1} \cdot {\mathbf{E}^{\mathcal{I}}}^{\top}, \quad
    \textbf{s}^{f}_{t} = \mathbf{o}^{f}_{t+1} \cdot {\mathbf{E}^{\mathcal{I}}}^{\top},
\end{equation}
where $\mathbf{E}^{\mathcal{I}} \in \mathbb{R}^{\mid \mathcal{I} \mid \times d}$ is the embedding matrix for all candidate items which reuses the input item embedding matrix to alleviate overfitting and reduce the model size.
Then we apply a softmax function to get the output distribution over candidate items:
\begin{equation}
  \hat{\textbf{y}}^{p}_{t} = softmax(\textbf{s}^{p}_{t}), \quad
  \hat{\textbf{y}}^{f}_{t} = softmax(\textbf{s}^{f}_{t}),
\end{equation}
where $\hat{\textbf{y}}^{p}_{t}, \hat{\textbf{y}}^{f}_{t} \in \mathbb{R}^{|\mathcal{I}|}$ denotes the estimated probabilities of each item to appear at position $t$ using past and future behavior information, respectively.

We employ the cross-entropy function to evaluate the prediction loss.
The final loss function is a combination of cross-entropy loss of two encoders as well as the regularization term:
\begin{equation}
\small
\begin{aligned}
\mathcal{L} &= \alpha \mathcal{L}_{past} + (1-\alpha)\mathcal{L}_{future} + \beta\mathcal{L}_{reg}\\
&= -\alpha\sum^{n-1}_{t =2}\text{OneHot}(i_{t}^{\star}) \log \hat{\textbf{y}}^{p}_{t}-(1-\alpha)\sum^{n-1}_{t=2}\text{OneHot}(i_{t}^{\star}) \log \hat{\textbf{y}}^{f}_{t} + \beta\mathcal{L}_{reg},
\end{aligned}
\end{equation}
where $i_{t}^{\star}$ is the ground truth item appears at position $t$, $\alpha \in [0,1]$ and $\beta > 0$ are two hyperparameters that control the contribution of dual tasks and the regularization term, respectively.

\subsubsection{Model Inference}
During inference, only the past encoder is used to make predictions based on users' historical interactions.
Since the input context of the past encoder is consistent between training and inference.
This will alleviate the training-inference gap problem in existing methods.

\subsection{Model Discussion}
\label{sec:model_discussion}
\subsubsection{Complexity Analysis}
This subsection analyzes the time and space complexity of our proposed dual network and bi-directional information transferring. 
The training time and space complexity of the dual network is determined by the backbone network used, i.e., twice the time and space complexity of the backbone.
While the inference time complexity of the dual network is the same as that of the backbone. 
For the Transformer-based backbone we use, both time and space complexity per layer of the dual network is $O(n^2d)$. 
Moreover, the time complexity of bi-directional information transferring is $O(h)$ as the multi-scale interest number is $h$; while its space complexity is $O(1)$ because it's a parameter-free operation.
Since $h<<n^2$, the time complexity of bi-directional information transferring can be ignored.
So the total time complexity of our proposed DualRec is approximate twice the complexity of the backbone, which is $O(n^2d)$ for our Transformer-based backbone.

\subsubsection{Compatibility Analysis}
The two core components of our proposed DualRec, namely the dual network and bi-directional information transferring, can be applied seamlessly to other sequential recommendation backbones, such as GRU4Rec, SASRec, FMLP-Rec and \textit{etc}. 
The dual network ensures past-future disentanglement while introducing future information in the training process; while bi-directional information transferring enables the mutual enhancement of past and future knowledge based on the dual network. 
These two components are generalized well to the sequential recommendation methods, which is demonstrated by the extensive compatibility analysis experiments conducted in Section \ref{sec:compatibility}. 
Notably, for the non-self-attention methods, multi-scale interest is captured by the multi-scale inputs.

%% file: experiments.tex
\section{Experiments}
We conduct extensive experiments on four real-world datasets to answer the following research questions:
\begin{itemize}
\item \textbf{(RQ1)} How does DualRec perform compared with state-of-the-art SR models? 
\item \textbf{(RQ2)} How do different components affect the performance of DualRec respectively? 
\item \textbf{(RQ3)} Can DualRec be compatible with mainstream SR models?
\item \textbf{(RQ4)} How do different hyperparameters affect DualRec?
\item \textbf{(RQ5)} How is future information used in DualRec?
\end{itemize}

\subsection{Experimental Setting}
\subsubsection{Datasets}
To comprehensively investigate the performance of DualRec, we conduct experiments on four large-scale real-world recommendation datasets:
\begin{itemize}[leftmargin=*]
    \item \textbf{Amazon Beauty, Sports, and Toys}\footnote{http://jmcauley.ucsd.edu/data/amazon/}:
    this series of datasets was crawled from relevant product reviews on Amazon.com \cite{McAuley2015Image}, which are widely used to evaluate recommendation performance \cite{kang2018self,zhou2020s3,zhou2022filter}. We select three categories: "Beauty", "Sports and Outdoors", and "Toys and Games" for our experiments.
    \item \textbf{Yelp}\footnote{https://www.yelp.com/dataset}:
    this is a dataset for the business recommendation, and we only use the user transaction records after \textit{January $1^{st}$, 2019} for our experiments.
\end{itemize}

For a fair comparison, we strictly follow the dataset settings of FMLP-Rec \cite{zhou2022filter}. 
We use the same pre-processed datasets and negative samples with FMLP-Rec.
Concretely, for all datasets, the user transaction records are grouped by userIDs and sorted by the transaction timestamps ascendingly to form the user behavioral sequence.
Moreover, the inactive users and unpopular items that have only fewer than five transaction records are filtered, following \cite{Bal2016Parallel,2010Factorizing}. 
Table \ref{tab:dataset} summarizes the statistics of the four datasets.

\begin{table}[ht]
\caption{Dataset Statistics}
\begin{tabular}{@{}c|c|c|c|c@{}}
\toprule
Dataset & \# Users & \# Items & \# Interactions & Density \\ \midrule
Beauty & 22,363 & 12,101 & 198,502 & 0.07\% \\
Sports & 25,598 & 18,357 & 296,337 & 0.05\% \\
Toys & 19,412 & 11,924 & 167,597 & 0.07\% \\
Yelp & 30,431 & 20,033 & 316,354 & 0.05\% \\
\bottomrule
\end{tabular}
\label{tab:dataset}
\end{table}

\subsubsection{Evaluation Metrics}
We evaluate models in terms of the top-K recommendation performance with three metrics, i.e., the Hit Ratio (HR@K), the Normalized Discounted Cumulative Gain (NDCG@K), and Mean Reciprocal Rank (MRR), which are widely used in related works \cite{kang2018self,sun2019bert4rec,zhou2022filter}.
HR@K is a recall-based metric that measures the average proportion of right items in the top-K recommendation lists.
NDCG@K evaluates the ranking quality of the top-K recommendation lists in a position-wise manner.
Specifically, we report the evaluation results on HR@\{1,5,10\}, NDCG@\{5,10\} and MRR, where we omit the NDCG@1 metric as it is equivalent to HR@1.
To make a fair comparison, we use the same negative samples as the FMLP-Rec \cite{zhou2022filter}.
For any ground-truth item, the 99 negative samples are selected randomly from the items that the user has not interacted with. 

\begin{table*}[!t]
\caption{Performance comparison using different methods on four popular sequential datasets. The best performance and the second-best performance methods are denoted in bold and underlined fonts respectively. The "*" mark denotes the statistical significance (p < 0.05) of comparing DualRec with the strongest baseline results and the “Improv.” column represents the relative improvement of DualRec over the strongest baseline. }
\centering
\setlength{\abovecaptionskip}{2mm}
\setlength{\belowcaptionskip}{-2mm}
\small
\setlength{\tabcolsep}{1mm}{
\begin{tabular}{llccccccccccccr}

\toprule
Datasets & Metric & GRU4Rec & Caser & HGN & RepeatNet & CLEA & SASRec & $\text{S3-Rec}_{\text{MIP}}$ & BERT4Rec & SRGNN & GCSAN & FMLP-Rec & DualRec & Improv.\\ \midrule
\multirow{6}{*}{Beauty} & HR@1   & 0.1519 & 0.1337 & 0.1683 & 0.1578 & 0.1325 & 0.1907 & 0.1678 & 0.1531 & 0.1729 & 0.1973 & \underline{0.2051} & $\textbf{0.2289}^*$ & +11.60\% \\
                        & HR@5   & 0.3612 & 0.3032 & 0.3544 & 0.3268 & 0.3305 & 0.4036 & 0.3710 & 0.3640 & 0.3518 & 0.3678 & \underline{0.4103} & $\textbf{0.4241}^*$ & +3.24\% \\
                        & NDCG5  & 0.2608 & 0.2219 & 0.2656 & 0.2455 & 0.2353 & 0.3022 & 0.2735 & 0.2622 & 0.2660 & 0.2864 & \underline{0.3133} & $\textbf{0.3320}^*$ & +5.97\% \\
                        & HR@10  & 0.4657 & 0.3942 & 0.4503 & 0.4205 & 0.4426 & 0.5043 & 0.4749 & 0.4739 & 0.4484 & 0.4542 & \underline{0.5070} & $\textbf{0.5190}^*$ & +2.37\%\\
                        & NDCG@10& 0.2944 & 0.2512 & 0.2965 & 0.2757 & 0.2715 & 0.3358 & 0.3069 & 0.2975 & 0.2971 & 0.3143 & \underline{0.3443} & $\textbf{0.3626}^*$ & +5.32\%\\
                        & MRR    & 0.2593 & 0.2263 & 0.2669 & 0.2498 & 0.2376 & 0.2990 & 0.2731 & 0.2614 & 0.2686 & 0.2882 & \underline{0.3102} & $\textbf{0.3302}^*$ & +6.45\%\\ \midrule
\multirow{6}{*}{Sports} & HR@1   & 0.1366 & 0.1135 & 0.1428 & 0.1334 & 0.1114 & 0.1676 & 0.1107 & 0.1255 & 0.1419 & 0.1669 & \underline{0.1722} & $\textbf{0.1947}^*$ & +13.07\%\\
                        & HR@5   & 0.3552 & 0.2866 & 0.3349 & 0.3162 & 0.3041 & \underline{0.3919} & 0.3141 & 0.3375 & 0.3367 & 0.3588 & 0.3886 & $\textbf{0.4127}^*$ & +5.31\%\\
                        & NDCG5  & 0.2487 & 0.2020 & 0.2420 & 0.2274 & 0.2096 & 0.2823 & 0.2143 & 0.2341 & 0.2418 & 0.2658 & \underline{0.2839} & $\textbf{0.3080}^*$ & +8.49\%\\
                        & HR@10  & 0.4853 & 0.4014 & 0.4551 & 0.4324 & 0.4274 & \underline{0.5169} & 0.4491 & 0.4772 & 0.4545 & 0.4737 & 0.5098 & $\textbf{0.5383}^*$ & +4.14\%\\
                        & NDCG@10 & 0.2907 & 0.2390 & 0.2806 & 0.2649 & 0.2493 & \underline{0.3244} & 0.2578 & 0.2775 & 0.2799 & 0.3029 & 0.3231 & $\textbf{0.3485}^*$ & +7.43\%\\
                        & MRR    & 0.2493 & 0.2100 & 0.2469 & 0.2334 & 0.2156 & \underline{0.2838} & 0.2203 & 0.2378 & 0.2461 & 0.2691 & 0.2830 & $\textbf{0.3078}^*$ & +8.47\%\\ \midrule
\multirow{6}{*}{Toys}   & HR@1   & 0.1303 & 0.1114 & 0.1504 & 0.1333 & 0.1104 & 0.1760 & 0.1825 & 0.1262 & 0.1600 & 0.1996 & \underline{0.2003} & $\textbf{0.2268}^*$ & +13.23\%\\
                        & HR@5   & 0.3526 & 0.2614 & 0.3276 & 0.3001 & 0.3055 & 0.3975 & 0.3892 & 0.3344 & 0.3389 & 0.3613 & \underline{0.4010} & $\textbf{0.4152}^*$ & +3.54\%\\
                        & NDCG5  & 0.2444 & 0.1885 & 0.2423 & 0.2192 & 0.2102 & 0.2907 & 0.2903 & 0.2327 & 0.2528 & 0.2836 & \underline{0.3055} & $\textbf{0.3253}^*$ & +6.48\%\\
                        & HR@10  & 0.4691 & 0.3540 & 0.4211 & 0.4015 & 0.4207 & \underline{0.5034} & 0.4935 & 0.4493 & 0.4413 & 0.4509 & 0.4977 & $\textbf{0.5145}^*$ & +2.21\%\\
                        & NDCG@10& 0.2820 & 0.2183 & 0.2724 & 0.2517 & 0.2473 & 0.3271 & 0.3239 & 0.2698 & 0.2857 & 0.3125 &\underline{0.3367} & $\textbf{0.3573}^*$ &+6.12\%\\
                        & MRR    & 0.2424 & 0.1967 & 0.2454 & 0.2253 & 0.2138 & 0.2877 & 0.2890 & 0.2338 & 0.2566 & 0.2871 & \underline{0.3034} & $\textbf{0.3256}^*$ &+7.32\%\\ \midrule
\multirow{6}{*}{Yelp}   & HR@1   & 0.1970 & 0.2188 & 0.2428 & 0.2341 & 0.2102 & 0.2327 & 0.2250 & 0.2405 & 0.2176 & 0.2493 & \underline{0.2625} & $\textbf{0.2893}^*$ & +10.21\%\\
                        & HR@5   & 0.5788 & 0.5111 & 0.5768 & 0.5357 & 0.5707 & 0.5949 & 0.5978 & 0.5976 & 0.5442 & 0.5725 & \underline{0.6246} & $\textbf{0.6328}^*$ & +1.32\%\\
                        & NDCG5  & 0.3933 & 0.3696 & 0.4162 & 0.3894 & 0.3955 & 0.4198 & 0.4171 & 0.4252 & 0.3860 & 0.4162 & \underline{0.4507} & $\textbf{0.4681}^*$ & +3.93\%\\
                        & HR@10  & 0.5788 & 0.6661 & 0.7411 & 0.6897 & 0.7473 & \underline{0.7722} & 0.7764 & 0.7597 & 0.7096 & 0.7371 & 0.7699 & $\textbf{0.7896}^*$ & +2.25\%\\
                        & NDCG@10  & 0.4511 & 0.4198 & 0.4695 & 0.4393 & 0.4527 & 0.4790 & 0.4751 & 0.4778 & 0.4395 & 0.4696 & \underline{0.4981} & $\textbf{0.5190}^*$ & +4.20\%\\
                        & MRR    & 0.3684 & 0.3595 & 0.3998 & 0.3769 & 0.3751 & 0.3994 & 0.3938 & 0.4026 & 0.3711 & 0.4006 & \underline{0.4236} & $\textbf{0.4466}^*$ & +5.43\%\\ \bottomrule  
\end{tabular}
\label{tab:comparison}}
\end{table*}

\subsubsection{Comparison Baselines}
To comprehensively demonstrate the effectiveness of the proposed DualRec framework, we compare it with eleven state-of-the-art recommendation baselines from different lines of research as listed below: 

\begin{itemize}
    \item Sequential recommendation models. (1) \textbf{Caser} \cite{tang2018personalized} captures short-term dynamic patterns of user activity through convolution. 
    (2) \textbf{GRU4Rec} \cite{hidasi2015session} uses the GRU to model sequences of user behavior. 
    (3) \textbf{HGN} \cite{ma2019hierarchical} captures long-term and short-term user interest through a hierarchical gating mechanism.
    (4) \textbf{RepeatNet} \cite{ren2019repeatnet} is based on an encoder-decoder architecture to select items from a user's transaction history records for repeat recommendations.
    (5) \textbf{SASRec} \cite{kang2018self} uses a causal transformer-based structure to predict the next item with the multi-head self-attention mechanism. 
    (6) \textbf{BERT4Rec} \cite{sun2019bert4rec} uses a bidirectional transformer model to train through a masked item prediction task.
    (7) \textbf{FMLP-Rec} \cite{zhou2022filter} is a all-MLP model that replaces the self-attention mechanism in the transformer structure with a learnable filter-enhanced block.
    \item Graph-based models. 
    (8) \textbf{SRGNN} \cite{wu2019session} uses graph attention networks to model session data as a session graph. 
    (9) \textbf{GCSAN} \cite{xu2019graph} combines graph neural networks with the self-attention mechanism for session recommendation.
    \item Contrastive learning augmented models. 
    (10) \textbf{S3-Rec} \cite{zhou2020s3} designed four pretraining tasks based on maximum mutual information to pretrain the self-attention-based sequential recommendation model. 
    We use only the Masked Item Prediction tasks without using attribute information for pretraining in our work. 
    (11) \textbf{CLEA} \cite{qin2021world} uses a contrastive learning approach to automatically extract items relevant to the target item to denoise the user behavior sequence. 
\end{itemize}

\subsubsection{Implementation Details}
We implement GRU4Rec, SASRec, S3-Rec, and FMLP-Rec using PyTorch \cite{paszke2017automatic}, and set hyperparameters according to the suggested values of their original work. 
For other methods (Caser, HGN, RepeatNet, CLEA, BERT4Rec, SRGNN and GCSAN), we follow the results of the reimplementations in FMLP-Rec \cite{zhou2022filter} and S3-Rec \cite{zhou2020s3}, as the datasets and evaluation metrics used in these works are strictly consistent with ours.

Our proposed DualRec is also implemented in PyTorch \cite{paszke2017automatic}. For a fair comparison, we set the maximum sequence length $n= 50$ and the dimensionality of the item embedding $d= 64$, which is consistent with \cite{zhou2022filter,zhou2020s3}. 
There are 2 Transformer layers and 8 heads per layer. 
Dropout is used to regularize the training process, and the dropout ratio is set to 0.5 for each dataset. 
We train with Adam optimizer \cite{kingma2014adam} with a learning rate of 0.001 and a weight decay of 1e-4. 
And the batch size is set as 256 for training, validation, and testing.

\subsection{Overall Performances (RQ1)}

\begin{table*}[!t]
\centering
\setlength{\abovecaptionskip}{2mm}
\setlength{\belowcaptionskip}{-2mm}
\caption{Ablation study results of different consisting components in DualRec.}
\begin{tabular}{ccccccccc}
\toprule
\multirow{2}{*}{Methods} & \multicolumn{2}{c}{Beauty} & \multicolumn{2}{c}{Sports} & \multicolumn{2}{c}{Toys} & \multicolumn{2}{c}{Yelp} \\
                        & HR@5   & NDCG@5 & HR@5   & NDCG@5 & HR@5   & NDCG@5 & HR@5   & NDCG@5 \\\midrule
DualRec               & \textbf{0.4241} & \textbf{0.3320} & \textbf{0.4127} & \textbf{0.3080} & \textbf{0.4152} & \textbf{0.3253} & \textbf{0.6328} & \textbf{0.4681} \\
DualRec w/o DE & 0.4154 & 0.3251 & 0.4037 & 0.3016 & 0.4098 & 0.3202 & 0.6273 & 0.4610 \\
DualRec w/o BIT       & 0.4235 & 0.3311 & 0.4102 & 0.3055 & 0.4122 & 0.3246 & 0.6284 & 0.4658 \\ \hline
DualRec w/o RPE       & 0.4216 & 0.3306 & 0.4093 & 0.3044 & 0.413  & 0.3234 & 0.6320 & 0.4677 \\
DualRec w/o LN        & 0.4086 & 0.3075 & 0.4097 & 0.3031 & 0.3987 & 0.2963 & 0.6321 & 0.4653 \\
DualRec w/o RC        & 0.3653 & 0.2642 & 0.3720 & 0.2632 & 0.3388 & 0.2360  & 0.5751 & 0.4052\\
\bottomrule
\end{tabular}
\label{tab:ablation}
\end{table*}

\begin{table*}[!ht]

\centering
\setlength{\abovecaptionskip}{2mm}
\setlength{\belowcaptionskip}{-2mm}
\caption{Compatibility Analysis with Different Backbones on four sequential datasets. Prefix "Dual" indicates the corresponding dual network model of backbone, and "BIT" means bi-directional information transferring. }
\setlength{\tabcolsep}{1mm}{
\begin{tabular}{l|ccc|ccc|ccc|ccc}
\toprule
Datasets      & \multicolumn{3}{c|}{Beauty} &\multicolumn{3}{c|}{Sports}&\multicolumn{3}{c|}{Toys}& \multicolumn{3}{c}{Yelp}\\
Model         &  HR@5   & NDCG@5 & MRR&  HR@5   & NDCG@5 & MRR & HR@5   & NDCG@5 & MRR  & HR@5   & NDCG@5 & MRR    \\ \midrule

SASRec        &  0.4036 & 0.3022 & 0.2990 & 0.3919 & 0.2823 & 0.2838 & 0.3975 & 0.2907 & 0.2877 & 0.5949 & 0.4198 & 0.3994\\
DualSASRec      & 0.4178 & 0.3147 & 0.3108 & 0.4055 & 0.2946 & 0.2936  & 0.4068 & 0.3038 & 0.3009 & 0.6198 & 0.4415 & 0.4183\\
DualSASRec+BIT   & \textbf{0.4223} & \textbf{0.3199} & \textbf{0.3153} & \textbf{0.4111} & \textbf{0.2985} & \textbf{0.2961}  & \textbf{0.4111} & \textbf{0.3084} & \textbf{0.3050}  & \textbf{0.6199} & \textbf{0.4457} & \textbf{0.4239} \\ \hline
GRU4Rec        & 0.3612 & 0.2608 & 0.2593 & 0.3552 & 0.2487 & 0.2493  & 0.3526 & 0.2444 & 0.2424& 0.5788 & 0.3933 & 0.3684 \\
DualGRU4Rec    & 0.3779 & 0.2726 & 0.2696  & 0.3681 & 0.2578 & 0.2575 & 0.3532 & 0.2454 & 0.2443 & 0.5869 & 0.4036 & 0.3803 \\
DualGRU4Rec+BIT  & \textbf{0.3866} & \textbf{0.2826} & \textbf{0.2798} & \textbf{0.3768} & \textbf{0.2638} & \textbf{0.2624} & \textbf{0.3679} & \textbf{0.2567} & \textbf{0.2535}& \textbf{0.5988} & \textbf{0.4141} & \textbf{0.3888} \\ \hline
FMLP-Rec    & 0.4103 & 0.3133 & 0.3102 & 0.3886 & 0.2839 & 0.2830 & 0.4010 & 0.3055 & 0.3034 & 0.6246 & 0.4507 & 0.4236 \\
DualFMLP-Rec   & 0.4172 & 0.3200 & 0.3170& 0.3937 & 0.2889 & 0.2878 & 0.4041 & 0.3084 & 0.3063 &  \textbf{0.6351} & \textbf{0.4708} & \textbf{0.4454} \\
DualFMLP-Rec+BIT & \textbf{0.4210} & \textbf{0.3240} & \textbf{0.3208} & \textbf{0.4002} & \textbf{0.2940} & \textbf{0.2914} & \textbf{0.4067} & \textbf{0.3109} & \textbf{0.3087}& 0.6343 & 0.4668 & 0.4407 \\ \bottomrule

\end{tabular}}
\label{tab:compatibility}
\end{table*}

The overall performance of the proposed DualRec model and baselines are presented in Table \ref{tab:comparison}. 
Based on the results, we summarize the following observations:
\begin{itemize}
    \item First, the Transformer-based methods (i.e., SASRec, S3-Rec and BERT4Rec) and the GNN-based methods (i.e., SRGNN and GCSAN) achieve better results than RNN-based and CNN-based models generally. 
    For Transformer-based methods, it can be attributed to the self-attention mechanism that better captures long-range dependencies in user interaction sequences. 
    For GNN-based methods, the great performance is because GNNs can better capture complex interactions of items than the conventional sequential recommendation method.
    \item Second, the methods using the masked item prediction (MIP) task for training (i.e., BERT4Rec)  or pretraining (i.e., S3-Rec) do not yield better results than SASRec. 
    These works are improvements on SASRec with the addition of MIP tasks to introduce future information, which should result in better performance than SASRec. 
    But the strong training-inference gap existing in MIP tasks (as described in Section \ref{sec:intro}) severely affects performance gains and may even cause performance degradation.

\item Finally, DualRec consistently achieves the best performances on all four datasets under all six evaluation metrics, demonstrating the superiority of our proposed method. 
It gains 12.03\% HR@1, 2.74\% HR@10, 5.77\% NDCG@10, and 6.92\% MRR improvements on average against the most competitive baselines.
The excellent performance of DualRec demonstrates the great utility of introducing future information into the training process.
The dual network structure in DualRec effectively avoids the training-inference gap in the MIP task by disentangling past and future information.
In Section \ref{sec:compatibility}, further extensive compatibility analysis experiments on other backbones are carried out to verify the high utility of our proposed framework for modeling future contexts. 
\end{itemize}

\subsection{Ablation Study (RQ2)}

As introduced in Section \ref{sec:method}, the backbone architecture of DualRec is constructed with two main components, i.e., the dual encoders (DE) and the bi-directional information transferring (BIT) component.
To verify the effectiveness of both components, we conduct the ablation study by removing either component from the DualRec architecture, which results in two sub-models:

\begin{itemize}

    \item \textit{DualRec w/o DE}: Without dual encoders (DE) means, the future encoder is removed. 
    So we utilize the past and future information to generate the corresponding representations on the same encoder with the left-to-right and right-to-right attention masks, respectively. 
    Then the past and future representations generated by the same encoder perform bidirectional information transferring between themselves.
    
    \item \textit{DualRec w/o BIT}: Without bi-directional information transferring, the shared embedding layer becomes the only associated component between the two encoders in the dual network.
\end{itemize}

Moreover, we also discuss other relevant components related to Transformer-based backbone structures, including relative positional embedding (\textit{DualRec w/o RPE}), layer normalization (\textit{DualRec w/o LN}), and the residual connection (\textit{DualRec w/o RC}).

The comparative results between DualRec and the sub-models are shown in Table \ref{tab:ablation}, where only the HR@10 and NDCG@10 scores are reported due to space limitation, from which we draw the following conclusions:
\begin{itemize}
    \item It can be observed that removing either component leads to performance degradation, proving that all components of the proposed model are effective and necessary. 
    Both the introduction of future information and the bi-directional information transferring positively contribute to the performance.
    What is more, the removal of residual connections in backbones (\textit{DualRec w/o RC}) leads to a network degradation problem that severely impairs the performance.
    
    \item The dual network model plays a more critical role in introducing future information into the training process, with bi-directional information transferring playing a relatively supportive role. 
    
\end{itemize}

\subsection{Compatibility Analysis (RQ3)}
\label{sec:compatibility}

DualRec is not only a specific model, but also a general framework that can be applied to most existing SR models.
To verify our claim, we replace the backbone with other representative SR models and evaluate the resulting models by comparing them with the original backbones.
The chosen SR models include Transformer-based SASRec, RNN-based GRU4Rec and Filter-based FMLP-Rec, and the comparison results are shown in Table \ref{tab:compatibility}. 

As we can see, for each backbone model, using the dual network to introduce future information into the training process achieves a consistent performance improvement on all four datasets. 
This performance improvement is due to the shared embedding layer learning knowledge from both past and future behaviors, while past-future disentanglement avoids severe training-inference gaps.

\begin{figure}[!ht]
    \centering
    \setlength{\abovecaptionskip}{2mm}
    \setlength{\belowcaptionskip}{-5mm}
    \begin{subfigure}{0.236\textwidth}
        \centering
        \setlength{\abovecaptionskip}{0cm}
    	\setlength{\belowcaptionskip}{-2mm}
        \includegraphics[width=1.68in]{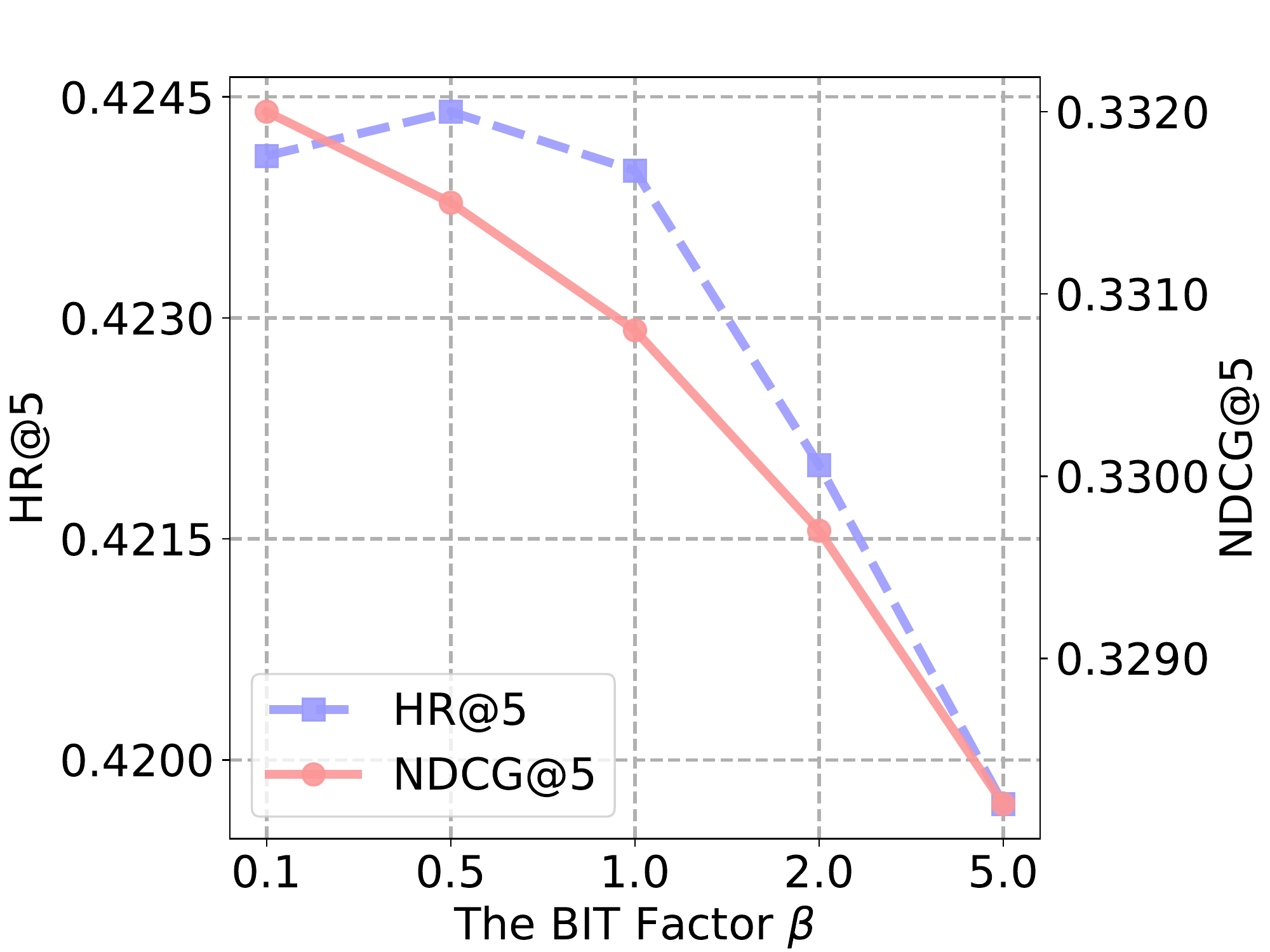} 
        \caption{Beauty}
    \end{subfigure}
    \begin{subfigure}{0.236\textwidth}
        \centering
        \setlength{\abovecaptionskip}{0cm}
    	\setlength{\belowcaptionskip}{-2mm}
        \includegraphics[width=1.68in]{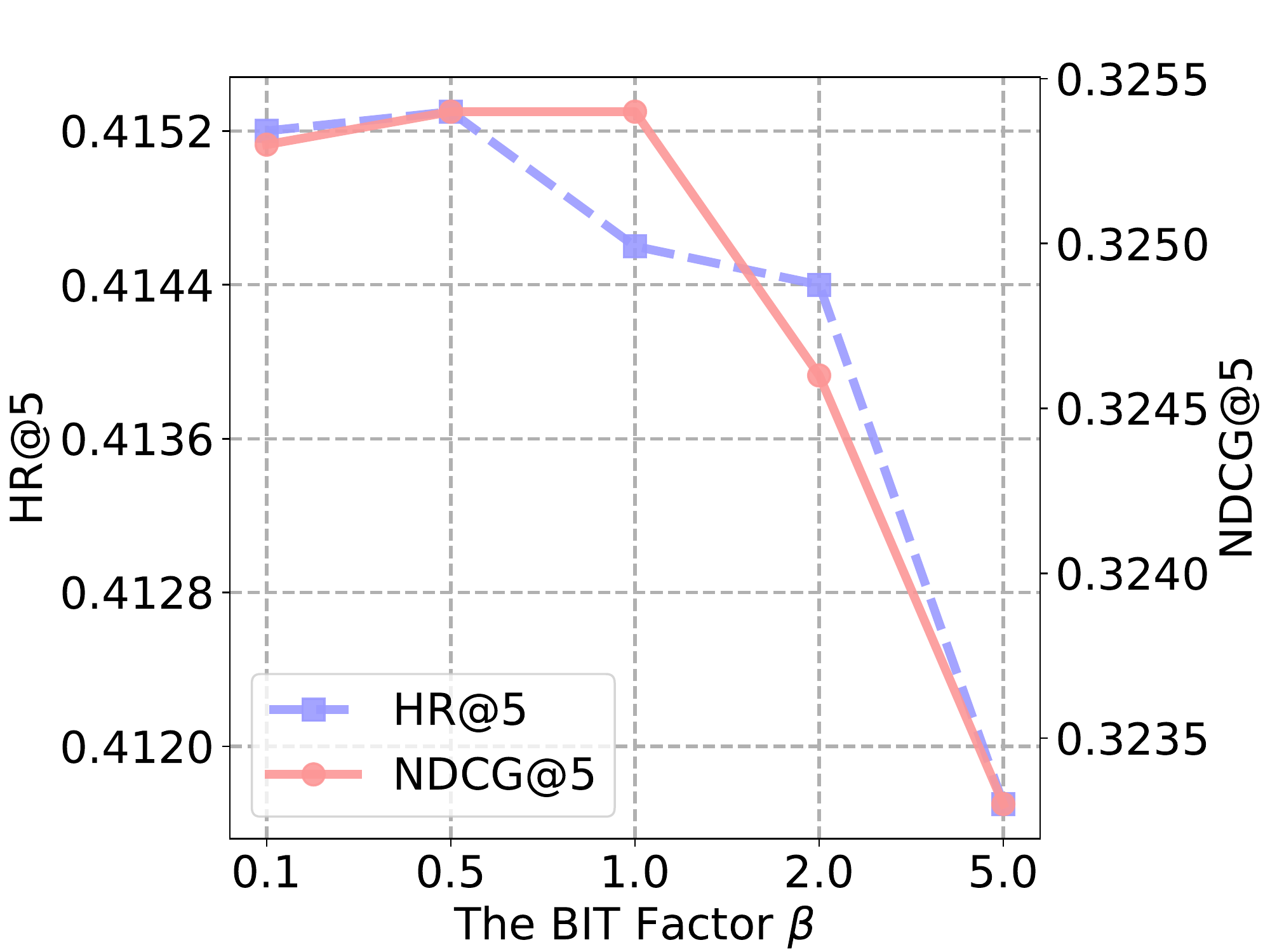} 
        \caption{Toys}
    \end{subfigure}
    \caption{Performances with different factor of bi-directional information transferring loss.}
    \label{fig:hyper4}
    \end{figure}

Based on the dual network, further enhancing the interaction between the past and future information using bi-directional information transferring can also facilitate improvement on the next item prediction training task. 
It is because information transferring between past and future promotes mutual learning and useful knowledge enhancement between the two encoders, thus improving the performance of the method. 
Furthermore, it is obvious that bi-directional knowledge transferring on DualFMLP-Rec causes performance degradation. 
This may be caused by the loss of original information along with the exchange of knowledge between the past and the future, and therefore the extent of knowledge transferring should not be too high which is also demonstrated by Figure \ref{fig:hyper4}. 

The above results demonstrate the broad generality and compatibility of DualRec on different backbone architectures, which can provide significant performance improvements to the backbone model.

 \begin{figure}[!ht]
    \centering
    \setlength{\abovecaptionskip}{2mm}
    \setlength{\belowcaptionskip}{-5mm}
    \begin{subfigure}{0.236\textwidth}
        \centering
        \setlength{\abovecaptionskip}{0cm}
    	\setlength{\belowcaptionskip}{-2mm}
        \includegraphics[width=1.68in]{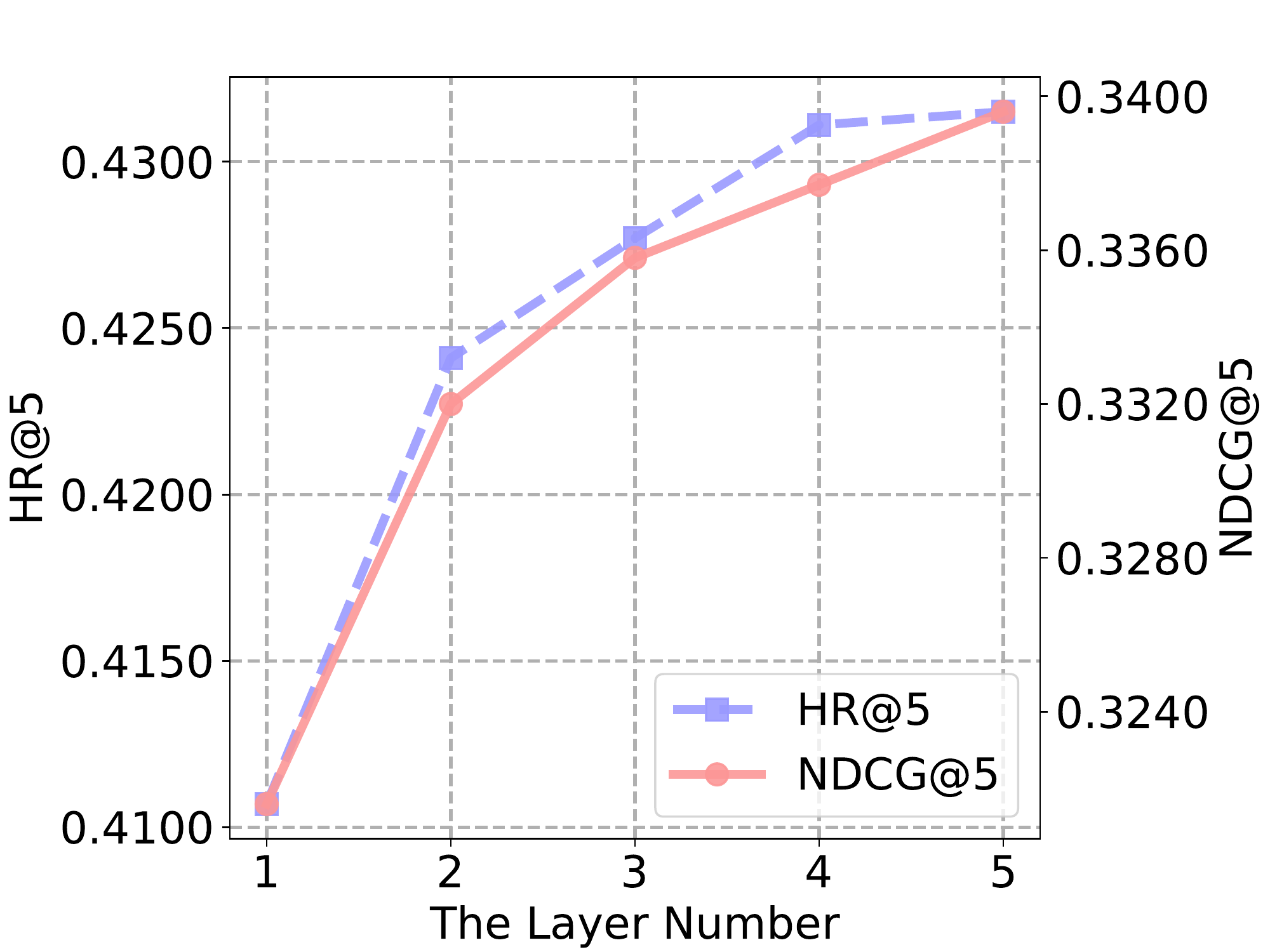} 
        \caption{Beauty}
    \end{subfigure}
    \begin{subfigure}{0.236\textwidth}
        \centering
        \setlength{\abovecaptionskip}{0cm}
    	\setlength{\belowcaptionskip}{-2mm}
        \includegraphics[width=1.68in]{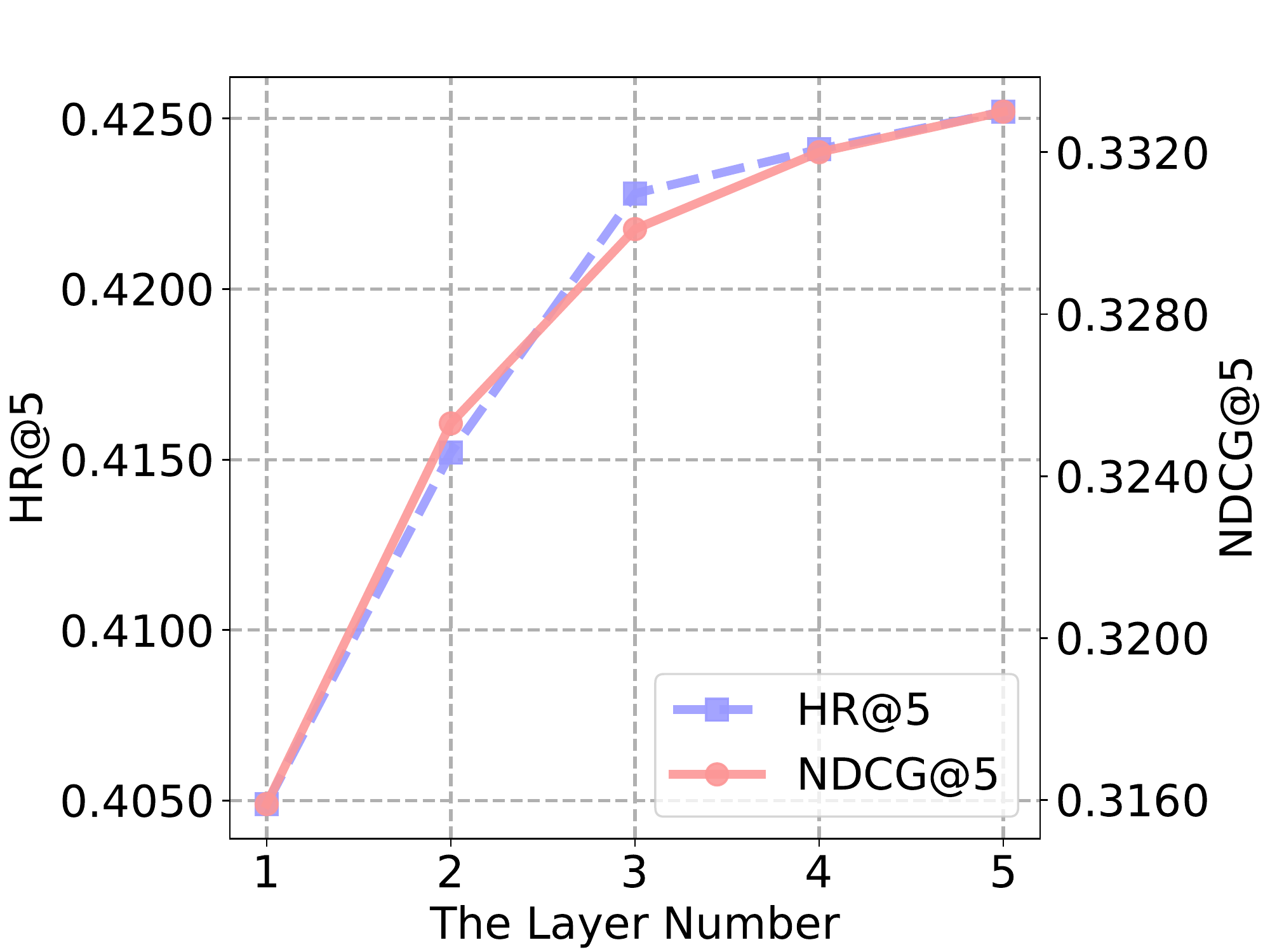} 
        \caption{Toys}
    \end{subfigure}
    \caption{Performances with different layer number.}
    \label{fig:hyper1}
    \end{figure}
    \begin{figure}[!ht]
    \centering
    \setlength{\abovecaptionskip}{2mm}
    \setlength{\belowcaptionskip}{-5mm}
    \begin{subfigure}{0.236\textwidth}
        \centering
        \setlength{\abovecaptionskip}{0cm}
    	\setlength{\belowcaptionskip}{-2mm}
        \includegraphics[width=1.68in]{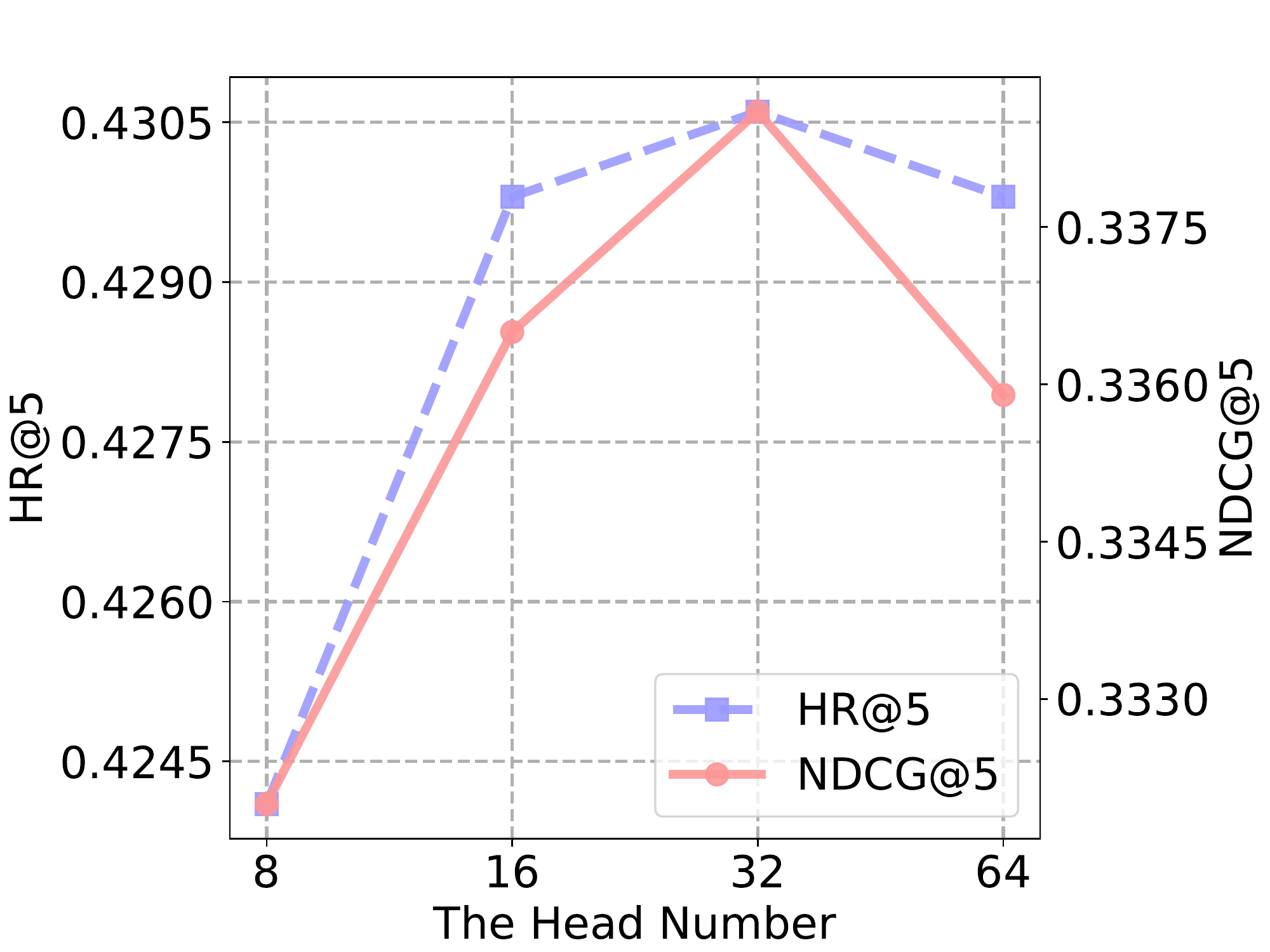} 
        \caption{Beauty}
    \end{subfigure}
    \begin{subfigure}{0.236\textwidth}
        \centering
        \setlength{\abovecaptionskip}{0cm}
    	\setlength{\belowcaptionskip}{-2mm}
        \includegraphics[width=1.68in]{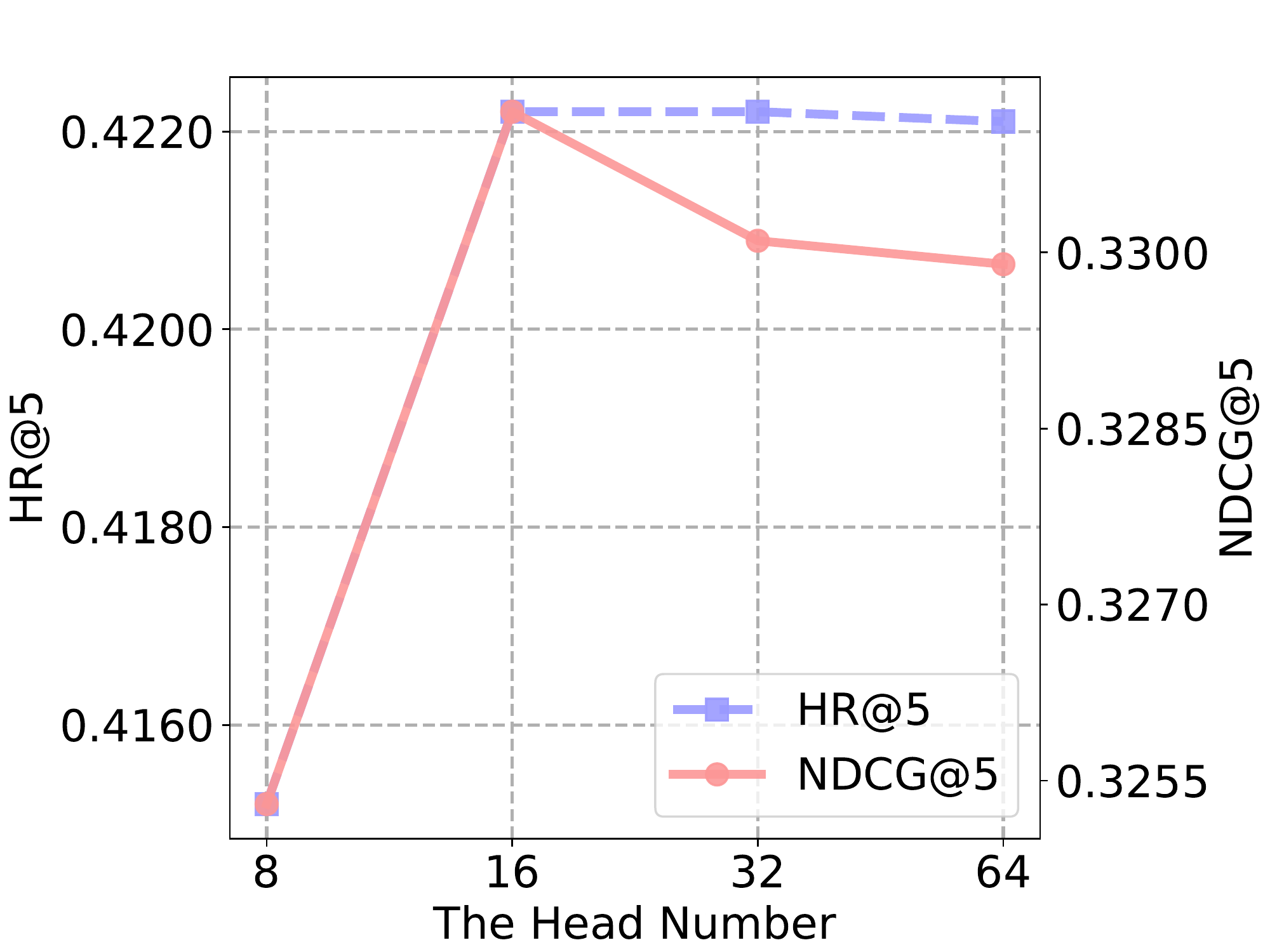} 
        \caption{Toys}
    \end{subfigure}
    \caption{Performances with different attention head number.}
    \label{fig:hyper2}
    \end{figure}
    
    \begin{figure}[!ht]
    \centering
    \setlength{\abovecaptionskip}{2mm}
    \setlength{\belowcaptionskip}{-5mm}
    \begin{subfigure}{0.236\textwidth}
        \centering
        \setlength{\abovecaptionskip}{0cm}
    	\setlength{\belowcaptionskip}{-2mm}
        \includegraphics[width=1.68in]{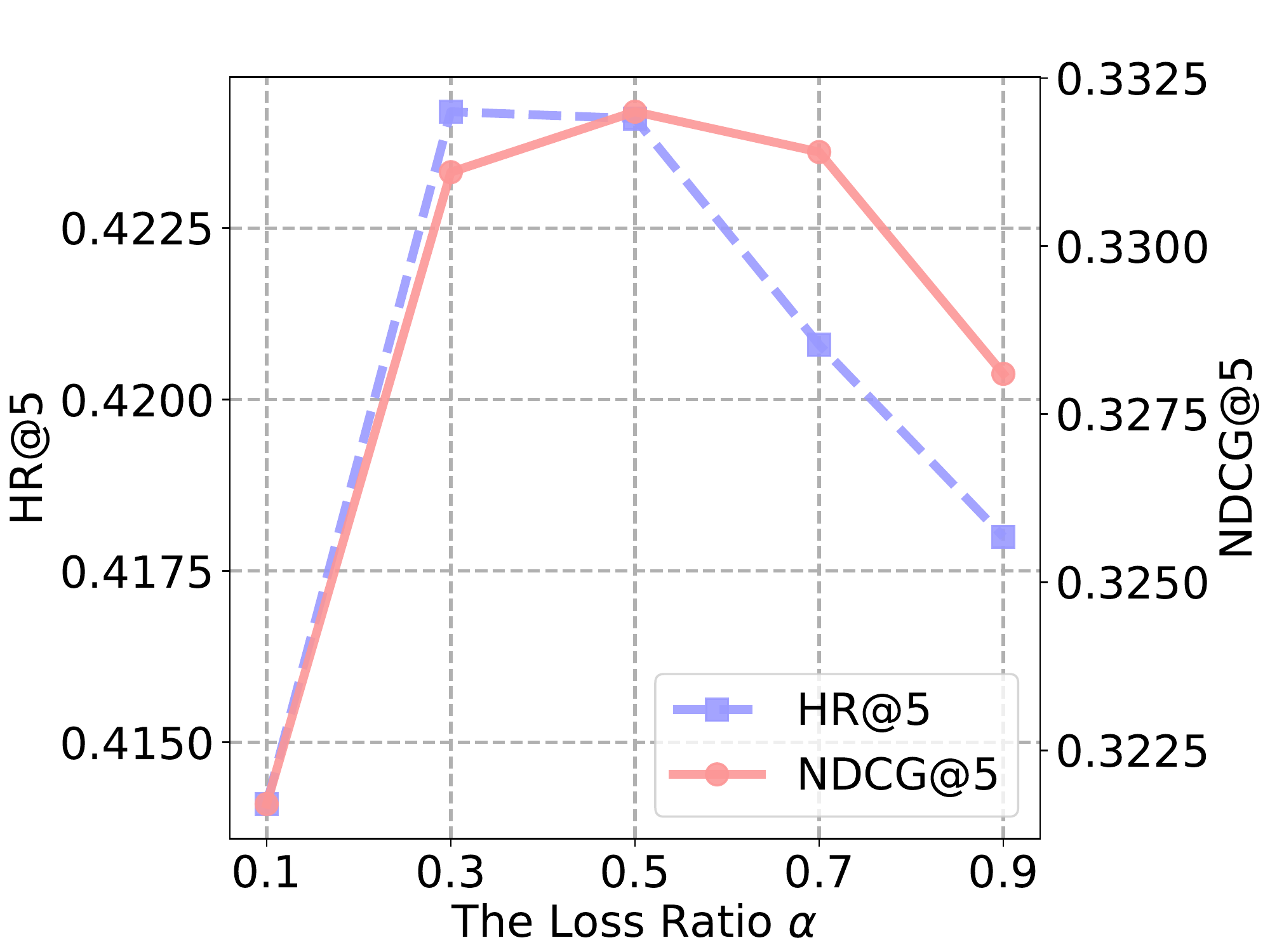} 
        \caption{Beauty}
    \end{subfigure}
    \begin{subfigure}{0.236\textwidth}
        \centering
        \setlength{\abovecaptionskip}{0cm}
    	\setlength{\belowcaptionskip}{-2mm}
        \includegraphics[width=1.68in]{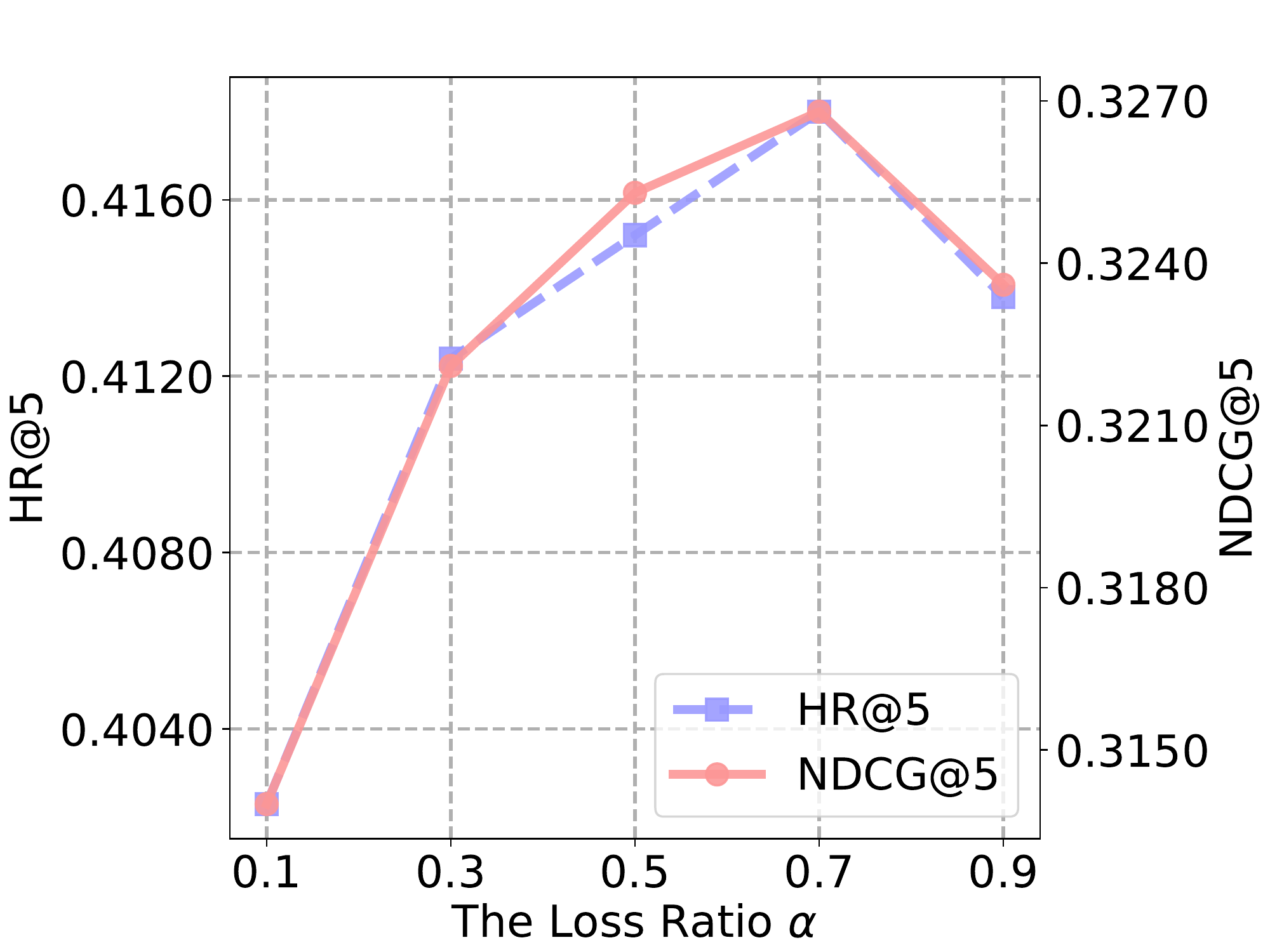} 
        \caption{Toys}
    \end{subfigure}
        \caption{Performances with different loss ratio of dual network model.}
    \label{fig:hyper3}
    \end{figure}
    
\subsection{Hyperparameter Analysis (RQ4)}
To investigate the impact of different hyperparameters on DualRec, we perform experiments on Beauty and Toys with different configurations of key hyperparameters, including (1) the number of layers $L$, (2) the number of self-attention heads $h$, (3) the loss ratio $\alpha$ of the dual network, and (4) the regularization factor $\beta$ of bi-directional information transferring.
We keep all other hyperparameters fixed when investigating the hyperparameter of interest.

\begin{itemize}
    \item \textbf{The number of layers $L$ and the number of attention heads $h$.} The number of layers and heads together determine the capacity of the base encoder, thus affecting the performance of the sequential representation learning. 
    Figure \ref{fig:hyper1} and \ref{fig:hyper2} show the performance with different number of layers $L$ and attention heads $h$, respectively.
    As $L$ or $h$ increases, i.e., the capacity of the model increases, the performance of the model also improves. Nevertheless, as they grow to a certain point, model performance tends to stabilize.
    The reason is that model with a large capacity tends to overfit noise in data, while model with a small capacity is not sufficient to capture user interest well. 

    \item \textbf{The loss ratio $\alpha$ of the dual network.} The loss ratio of the dual network defines the weight of the prediction task loss for the past encoder and the future encoder. 
    It controls the ratio of past and future information utilized in the dual network model. 
    Overall, the best results are obtained when the losses in the two encoders of the dual network model are relatively balanced, i.e., when the loss ratio is around 0.5, as shown in Figure \ref{fig:hyper3}. 
    Because either too small or too large $\alpha$ will cause the past or future encoder to be undertrained. 
    Therefore, keeping the weights of training loss approximately equal between the two encoders is beneficial for stable training.

    \item \textbf{The factor of bi-directional information transferring loss $\beta$.} The factor $\beta$ determines the extent of bi-directional information transferring, i.e., the coefficient of the regular term. Figure \ref{fig:hyper4} shows that the factor $\beta$ should not be set too high,  which would affect the performance of the model.
    This performance degradation may result from excessive $\beta$ leading to loss of original information during bidirectional information transferring.
    At the same time, if the factor $\beta$ is set too small, knowledge transfer will not work effectively, so $\beta$ can be set to a moderate value, such as 0.5 in the Figure \ref{fig:hyper4}.

\end{itemize}

\subsection{Future Information Utilization (RQ5)}

We delve deep into how future information is used in DualRec.
First, as sequential dependencies of user behaviors may not be strictly held, we can train SR models in different ways, including (1) \textbf{Past-Only Model}: standard next-item-prediction with only past behaviors, (2) \textbf{Future-Only Model}: previous-item prediction with only future behaviors, and (3) \textbf{DualRec}: dual tasks with both past and future behaviors.
But when testing, we follow the standard setup of predicting users' next interactions based on their historical interactions.
Table \ref{tab:effectiveness} shows the results of these three settings.
Not surprisingly, due to the extreme case of training-inference gap, training with only future data performs worse than training with past data.
But it is acceptable, for example, the performance only drops by 8.0\% compared to using only past information in terms of HR@5 on the Beauty dataset (0.3833 and 0.4166, respectively).
It suggests that the knowledge learned in the future data is beneficial for the next-item-prediction task. 
And this is why combining past and future data yields better results.

\begin{table}[!t]
\centering 
\setlength{\abovecaptionskip}{2mm}
\setlength{\belowcaptionskip}{-2mm}
\caption{The performance of Past-Only Model, Future-Only Model, and DualRec on four datasets. }
\begin{tabular}{llccc}
\toprule
Datasets                & Metric & Past-Only & Future-Only & DualRec \\\midrule
\multirow{2}{*}{Beauty} & HR@5   & 0.4166    & 0.3833      & \textbf{0.4235}         \\
                        & NDCG@5 & 0.3272    & 0.2912      & \textbf{0.3311}         \\
\multirow{2}{*}{Sports} & HR@5   & 0.4061    & 0.3786      & \textbf{0.4102}         \\
                        & NDCG@5 & 0.3027    & 0.2766      & \textbf{0.3055}         \\
\multirow{2}{*}{Toys}   & HR@5   & 0.4085    & 0.3601      & \textbf{0.4122}         \\
                        & NDCG@5 & 0.3212    & 0.2761      & \textbf{0.3246}         \\
\multirow{2}{*}{Yelp}   & HR@5   & 0.6276    & 0.6215      & \textbf{0.6284}         \\
                        & NDCG@5 & 0.4628    & 0.4576      & \textbf{0.4658}        \\\bottomrule
\end{tabular}
\label{tab:effectiveness}
\end{table}

\begin{figure}[!t]
\centering
\setlength{\abovecaptionskip}{2mm}
\setlength{\belowcaptionskip}{-6mm}
\begin{subfigure}{0.236\textwidth}
    \centering
    \setlength{\abovecaptionskip}{0cm}
    	\setlength{\belowcaptionskip}{-2mm}
    \includegraphics[width=1.68in]{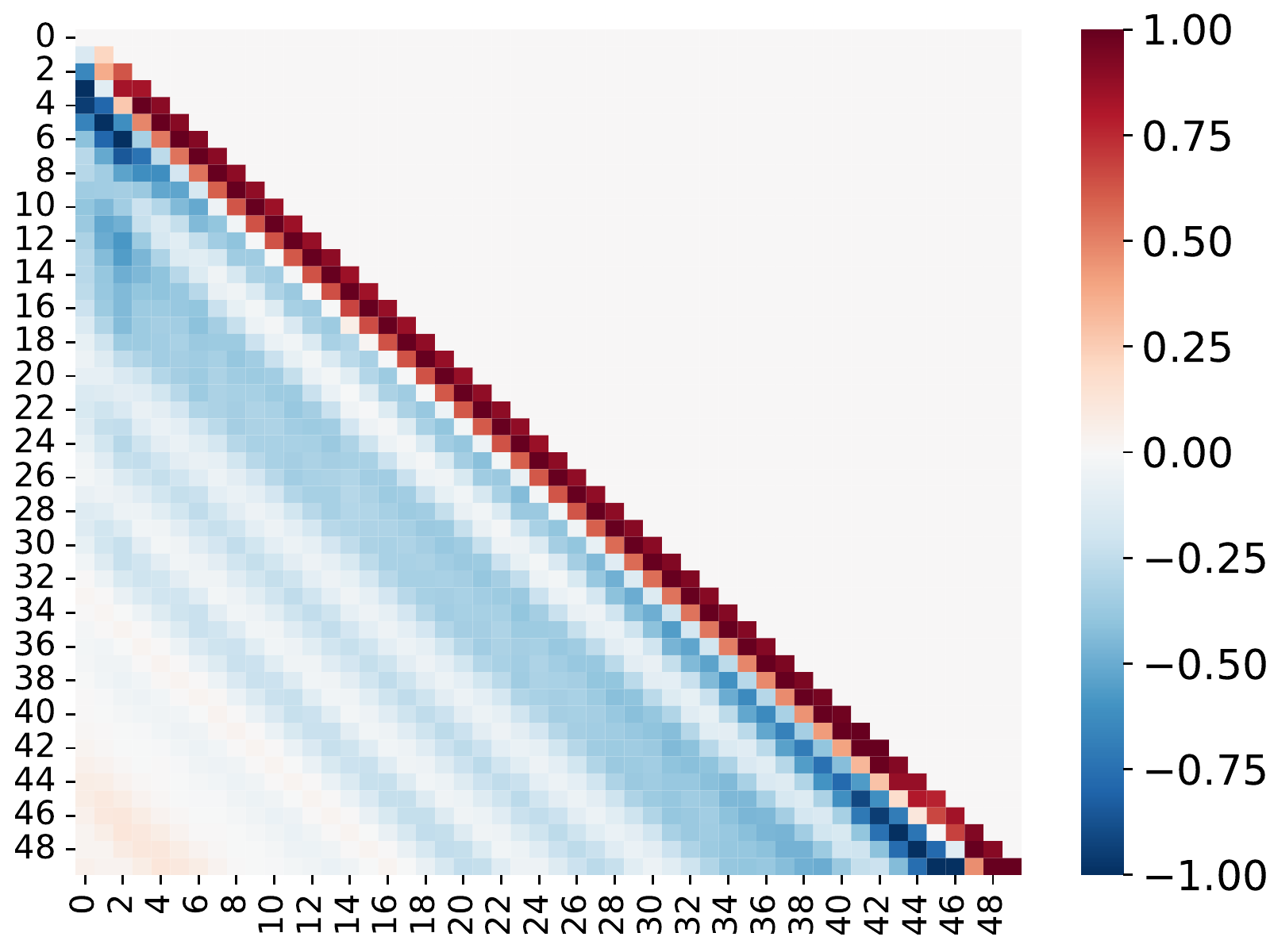} 
    \caption{Layer1}
\end{subfigure}
\begin{subfigure}{0.236\textwidth}
    \centering
    \setlength{\abovecaptionskip}{0cm}
    	\setlength{\belowcaptionskip}{-1mm}
    \includegraphics[width=1.68in]{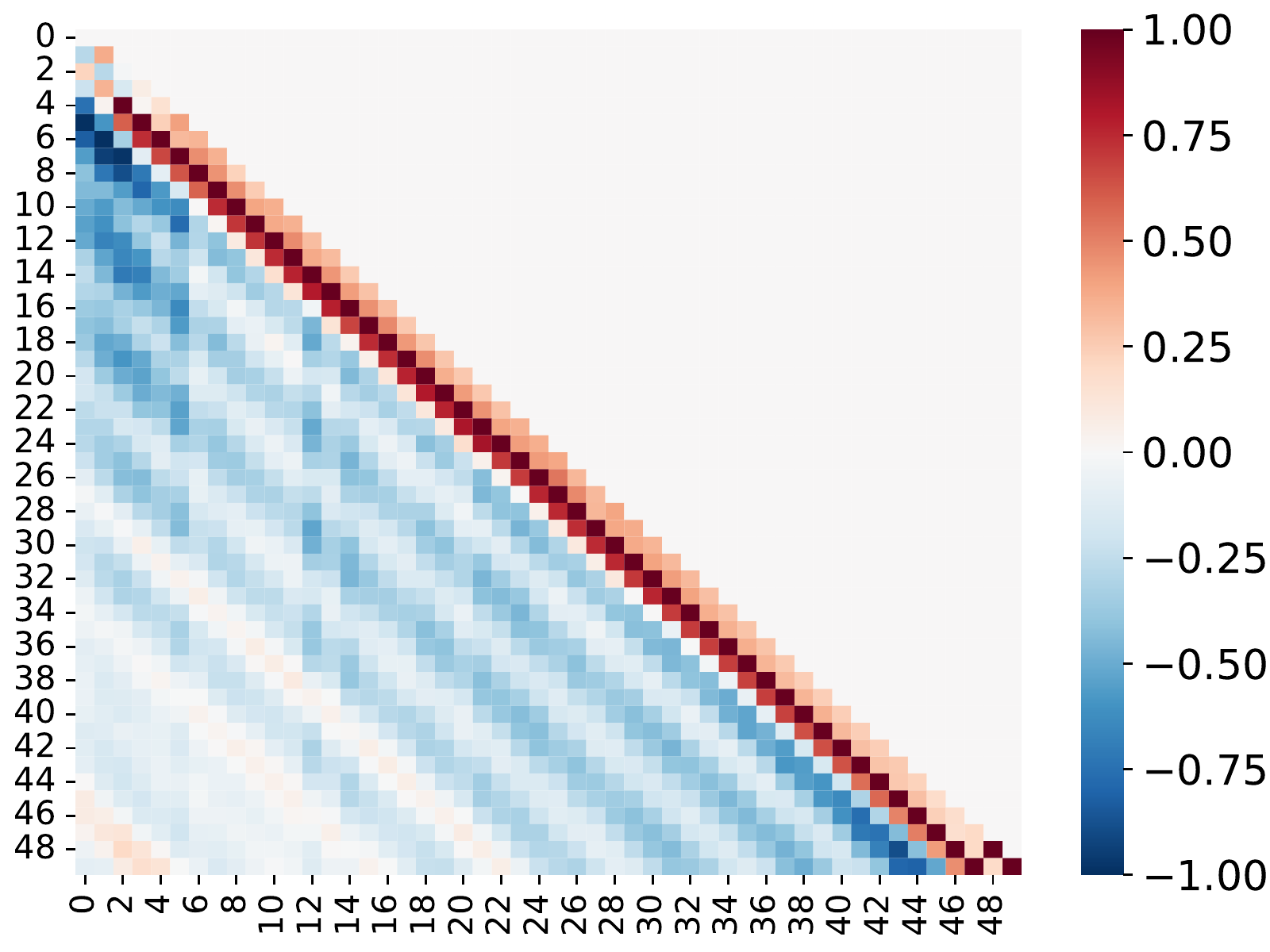} 
    \caption{Layer2}
\end{subfigure}
\caption{Heat maps of the average difference in attention score per layer between the past encoder in the DualRec and the Past-Only Model on the Yelp test dataset ({\color{red}red} means the average attention scores in the DualRec are higher than those in the Past-Only Model; and {\color{blue}blue} indicates the opposite case). The coordinates in the figure indicate the sequence index. }

\label{fig:avg}

\end{figure}

We further explore the impact of the future information.
We compare the past encoder in DualRec (DualRec-Past for simplicity) and the Past-Only Model, since they are structurally identical and differ only in training settings (past encoder in DualRec can leverage future information).
Specifically, we calculate the average difference in attention scores between the two models on the Yelp testing dataset.
The heat maps are shown in Figure \ref{fig:avg}, where red means the average attention scores in DualRec-Past are higher than those in the Past-Only Model, and blue indicates the opposite case. 
Noticed that we set the number of attention head as 1 for simplicity. 
From Figure \ref{fig:avg}, the attention scores of DualRec-Past are significantly higher around the diagonal than those of Past-Only Model (the diagonal line of the heat map appears red). 
That can attribute to the dual tasks introducing future information, where the self-attention mechanism focuses more on items that describe users' current preference, i.e., short-term items.
The widely recognized Markov property \cite{2010Factorizing} in sequential recommendation indicates that short-range history plays a more important role than long-range history in capturing users' real-time interest. 
So with the help of future information, DualRec-Past can better focus on the more valuable short-range history, resulting in significant performance improvements.

%% file: sigconf.bbl

\begin{thebibliography}{40}


\ifx \showCODEN    \undefined \def \showCODEN     #1{\unskip}     \fi
\ifx \showDOI      \undefined \def \showDOI       #1{#1}\fi
\ifx \showISBNx    \undefined \def \showISBNx     #1{\unskip}     \fi
\ifx \showISBNxiii \undefined \def \showISBNxiii  #1{\unskip}     \fi
\ifx \showISSN     \undefined \def \showISSN      #1{\unskip}     \fi
\ifx \showLCCN     \undefined \def \showLCCN      #1{\unskip}     \fi
\ifx \shownote     \undefined \def \shownote      #1{#1}          \fi
\ifx \showarticletitle \undefined \def \showarticletitle #1{#1}   \fi
\ifx \showURL      \undefined \def \showURL       {\relax}        \fi
\providecommand\bibfield[2]{#2}
\providecommand\bibinfo[2]{#2}
\providecommand\natexlab[1]{#1}
\providecommand\showeprint[2][]{arXiv:#2}

\bibitem[min(2020)]%
        {mindspore}
 \bibinfo{year}{2020}\natexlab{}.
\newblock \bibinfo{title}{MindSpore}.
\newblock
\newblock
\urldef\tempurl%
\url{https://www.mindspore.cn}
\showURL{%
\tempurl}


\bibitem[Ba et~al\mbox{.}(2016)]%
        {ba2016layer}
\bibfield{author}{\bibinfo{person}{Jimmy~Lei Ba}, \bibinfo{person}{Jamie~Ryan
  Kiros}, {and} \bibinfo{person}{Geoffrey~E Hinton}.}
  \bibinfo{year}{2016}\natexlab{}.
\newblock \showarticletitle{Layer normalization}.
\newblock \bibinfo{journal}{\emph{arXiv preprint arXiv:1607.06450}}
  (\bibinfo{year}{2016}).
\newblock


\bibitem[Cheng et~al\mbox{.}(2016)]%
        {wide-deep}
\bibfield{author}{\bibinfo{person}{Heng{-}Tze Cheng}, \bibinfo{person}{Levent
  Koc}, \bibinfo{person}{Jeremiah Harmsen}, \bibinfo{person}{Tal Shaked},
  \bibinfo{person}{Tushar Chandra}, \bibinfo{person}{Hrishi Aradhye},
  \bibinfo{person}{Glen Anderson}, \bibinfo{person}{Greg Corrado},
  \bibinfo{person}{Wei Chai}, \bibinfo{person}{Mustafa Ispir},
  \bibinfo{person}{Rohan Anil}, \bibinfo{person}{Zakaria Haque},
  \bibinfo{person}{Lichan Hong}, \bibinfo{person}{Vihan Jain},
  \bibinfo{person}{Xiaobing Liu}, {and} \bibinfo{person}{Hemal Shah}.}
  \bibinfo{year}{2016}\natexlab{}.
\newblock \showarticletitle{Wide {\&} Deep Learning for Recommender Systems}.
  In \bibinfo{booktitle}{\emph{Proc. Workshop Deep Learning for Recommender
  Systems}}.
\newblock


\bibitem[Davidson et~al\mbox{.}(2010)]%
        {davidson2010youtube}
\bibfield{author}{\bibinfo{person}{James Davidson}, \bibinfo{person}{Benjamin
  Liebald}, \bibinfo{person}{Junning Liu}, \bibinfo{person}{Palash Nandy},
  \bibinfo{person}{Taylor Van~Vleet}, \bibinfo{person}{Ullas Gargi},
  \bibinfo{person}{Sujoy Gupta}, \bibinfo{person}{Yu He}, \bibinfo{person}{Mike
  Lambert}, \bibinfo{person}{Blake Livingston}, {et~al\mbox{.}}}
  \bibinfo{year}{2010}\natexlab{}.
\newblock \showarticletitle{The YouTube video recommendation system}. In
  \bibinfo{booktitle}{\emph{Proceedings of the fourth ACM conference on
  Recommender systems}}. \bibinfo{pages}{293--296}.
\newblock


\bibitem[Deldjoo et~al\mbox{.}(2016)]%
        {deldjoo2016video}
\bibfield{author}{\bibinfo{person}{Yashar Deldjoo}, \bibinfo{person}{Mehdi
  Elahi}, \bibinfo{person}{Paolo Cremonesi}, \bibinfo{person}{Franca Garzotto},
  \bibinfo{person}{Pietro Piazzolla}, {and} \bibinfo{person}{Massimo
  Quadrana}.} \bibinfo{year}{2016}\natexlab{}.
\newblock \showarticletitle{Content-based video recommendation system based on
  stylistic visual features}.
\newblock \bibinfo{journal}{\emph{Journal on Data Semantics}}
  \bibinfo{volume}{5}, \bibinfo{number}{2} (\bibinfo{year}{2016}),
  \bibinfo{pages}{99--113}.
\newblock


\bibitem[Glorot et~al\mbox{.}(2011)]%
        {glorot2011deep}
\bibfield{author}{\bibinfo{person}{Xavier Glorot}, \bibinfo{person}{Antoine
  Bordes}, {and} \bibinfo{person}{Yoshua Bengio}.}
  \bibinfo{year}{2011}\natexlab{}.
\newblock \showarticletitle{Deep sparse rectifier neural networks}. In
  \bibinfo{booktitle}{\emph{Proceedings of the fourteenth international
  conference on artificial intelligence and statistics}}. JMLR Workshop and
  Conference Proceedings, \bibinfo{pages}{315--323}.
\newblock


\bibitem[Guo et~al\mbox{.}(2017)]%
        {deepfm}
\bibfield{author}{\bibinfo{person}{Huifeng Guo}, \bibinfo{person}{Ruiming
  Tang}, \bibinfo{person}{Yunming Ye}, \bibinfo{person}{Zhenguo Li}, {and}
  \bibinfo{person}{Xiuqiang He}.} \bibinfo{year}{2017}\natexlab{}.
\newblock \showarticletitle{DeepFM: {A} Factorization-Machine based Neural
  Network for {CTR} Prediction}. In \bibinfo{booktitle}{\emph{IJCAI}}.
\newblock


\bibitem[He et~al\mbox{.}(2016)]%
        {he2016deep}
\bibfield{author}{\bibinfo{person}{Kaiming He}, \bibinfo{person}{Xiangyu
  Zhang}, \bibinfo{person}{Shaoqing Ren}, {and} \bibinfo{person}{Jian Sun}.}
  \bibinfo{year}{2016}\natexlab{}.
\newblock \showarticletitle{Deep residual learning for image recognition}. In
  \bibinfo{booktitle}{\emph{Proceedings of the IEEE conference on computer
  vision and pattern recognition}}. \bibinfo{pages}{770--778}.
\newblock


\bibitem[Hidasi and Karatzoglou(2018)]%
        {hidasi2018recurrent}
\bibfield{author}{\bibinfo{person}{Bal{\'a}zs Hidasi} {and}
  \bibinfo{person}{Alexandros Karatzoglou}.} \bibinfo{year}{2018}\natexlab{}.
\newblock \showarticletitle{Recurrent neural networks with top-k gains for
  session-based recommendations}. In \bibinfo{booktitle}{\emph{Proceedings of
  the 27th ACM international conference on information and knowledge
  management}}. \bibinfo{pages}{843--852}.
\newblock


\bibitem[Hidasi et~al\mbox{.}(2015)]%
        {hidasi2015session}
\bibfield{author}{\bibinfo{person}{Bal{\'a}zs Hidasi},
  \bibinfo{person}{Alexandros Karatzoglou}, \bibinfo{person}{Linas Baltrunas},
  {and} \bibinfo{person}{Domonkos Tikk}.} \bibinfo{year}{2015}\natexlab{}.
\newblock \showarticletitle{Session-based recommendations with recurrent neural
  networks}.
\newblock \bibinfo{journal}{\emph{arXiv preprint arXiv:1511.06939}}
  (\bibinfo{year}{2015}).
\newblock


\bibitem[Hidasi et~al\mbox{.}(2016)]%
        {Bal2016Parallel}
\bibfield{author}{\bibinfo{person}{Balázs Hidasi}, \bibinfo{person}{Massimo
  Quadrana}, \bibinfo{person}{Alexandros Karatzoglou}, {and}
  \bibinfo{person}{Domonkos Tikk}.} \bibinfo{year}{2016}\natexlab{}.
\newblock \showarticletitle{Parallel Recurrent Neural Network Architectures for
  Feature-rich Session-based Recommendations}. In \bibinfo{booktitle}{\emph{Acm
  Conference on Recommender Systems}}.
\newblock


\bibitem[Hinton et~al\mbox{.}(2015)]%
        {hinton2015distilling}
\bibfield{author}{\bibinfo{person}{Geoffrey Hinton}, \bibinfo{person}{Oriol
  Vinyals}, \bibinfo{person}{Jeff Dean}, {et~al\mbox{.}}}
  \bibinfo{year}{2015}\natexlab{}.
\newblock \showarticletitle{Distilling the knowledge in a neural network}.
\newblock \bibinfo{journal}{\emph{arXiv preprint arXiv:1503.02531}}
  \bibinfo{volume}{2}, \bibinfo{number}{7} (\bibinfo{year}{2015}).
\newblock


\bibitem[Hu et~al\mbox{.}(2020)]%
        {hu2020graph}
\bibfield{author}{\bibinfo{person}{Linmei Hu}, \bibinfo{person}{Chen Li},
  \bibinfo{person}{Chuan Shi}, \bibinfo{person}{Cheng Yang}, {and}
  \bibinfo{person}{Chao Shao}.} \bibinfo{year}{2020}\natexlab{}.
\newblock \showarticletitle{Graph neural news recommendation with long-term and
  short-term interest modeling}.
\newblock \bibinfo{journal}{\emph{Information Processing \& Management}}
  \bibinfo{volume}{57}, \bibinfo{number}{2} (\bibinfo{year}{2020}),
  \bibinfo{pages}{102142}.
\newblock


\bibitem[Kang and McAuley(2018)]%
        {kang2018self}
\bibfield{author}{\bibinfo{person}{Wang-Cheng Kang} {and}
  \bibinfo{person}{Julian McAuley}.} \bibinfo{year}{2018}\natexlab{}.
\newblock \showarticletitle{Self-attentive sequential recommendation}. In
  \bibinfo{booktitle}{\emph{2018 IEEE International Conference on Data Mining
  (ICDM)}}. IEEE, \bibinfo{pages}{197--206}.
\newblock


\bibitem[Kingma and Ba(2014)]%
        {kingma2014adam}
\bibfield{author}{\bibinfo{person}{Diederik~P Kingma} {and}
  \bibinfo{person}{Jimmy Ba}.} \bibinfo{year}{2014}\natexlab{}.
\newblock \showarticletitle{Adam: A method for stochastic optimization}.
\newblock \bibinfo{journal}{\emph{arXiv preprint arXiv:1412.6980}}
  (\bibinfo{year}{2014}).
\newblock


\bibitem[Kipf and Welling(2016)]%
        {kipf2016gcn}
\bibfield{author}{\bibinfo{person}{Thomas~N Kipf} {and} \bibinfo{person}{Max
  Welling}.} \bibinfo{year}{2016}\natexlab{}.
\newblock \showarticletitle{Semi-supervised classification with graph
  convolutional networks}.
\newblock \bibinfo{journal}{\emph{arXiv preprint arXiv:1609.02907}}
  (\bibinfo{year}{2016}).
\newblock


\bibitem[Krizhevsky et~al\mbox{.}(2012)]%
        {krizhevsky2012imagenet}
\bibfield{author}{\bibinfo{person}{Alex Krizhevsky}, \bibinfo{person}{Ilya
  Sutskever}, {and} \bibinfo{person}{Geoffrey~E Hinton}.}
  \bibinfo{year}{2012}\natexlab{}.
\newblock \showarticletitle{Imagenet classification with deep convolutional
  neural networks}.
\newblock \bibinfo{journal}{\emph{Advances in neural information processing
  systems}}  \bibinfo{volume}{25} (\bibinfo{year}{2012}),
  \bibinfo{pages}{1097--1105}.
\newblock


\bibitem[Ma et~al\mbox{.}(2019)]%
        {ma2019hierarchical}
\bibfield{author}{\bibinfo{person}{Chen Ma}, \bibinfo{person}{Peng Kang}, {and}
  \bibinfo{person}{Xue Liu}.} \bibinfo{year}{2019}\natexlab{}.
\newblock \showarticletitle{Hierarchical gating networks for sequential
  recommendation}. In \bibinfo{booktitle}{\emph{Proceedings of the 25th ACM
  SIGKDD international conference on knowledge discovery \& data mining}}.
  \bibinfo{pages}{825--833}.
\newblock


\bibitem[McAuley et~al\mbox{.}(2015)]%
        {McAuley2015Image}
\bibfield{author}{\bibinfo{person}{Julian~J. McAuley},
  \bibinfo{person}{Christopher Targett}, \bibinfo{person}{Qinfeng Shi}, {and}
  \bibinfo{person}{Anton van~den Hengel}.} \bibinfo{year}{2015}\natexlab{}.
\newblock \showarticletitle{Image-Based Recommendations on Styles and
  Substitutes}. In \bibinfo{booktitle}{\emph{Proceedings of SIGIR}}.
  \bibinfo{publisher}{{ACM}}, \bibinfo{pages}{43--52}.
\newblock


\bibitem[Mikolov et~al\mbox{.}(2010)]%
        {mikolov2010recurrent}
\bibfield{author}{\bibinfo{person}{Tomas Mikolov}, \bibinfo{person}{Martin
  Karafi{\'a}t}, \bibinfo{person}{Lukas Burget}, \bibinfo{person}{Jan
  Cernock{\`y}}, {and} \bibinfo{person}{Sanjeev Khudanpur}.}
  \bibinfo{year}{2010}\natexlab{}.
\newblock \showarticletitle{Recurrent neural network based language model.}. In
  \bibinfo{booktitle}{\emph{Interspeech}}, Vol.~\bibinfo{volume}{2}. Makuhari,
  \bibinfo{pages}{1045--1048}.
\newblock


\bibitem[Paszke et~al\mbox{.}(2017)]%
        {paszke2017automatic}
\bibfield{author}{\bibinfo{person}{Adam Paszke}, \bibinfo{person}{Sam Gross},
  \bibinfo{person}{Soumith Chintala}, \bibinfo{person}{Gregory Chanan},
  \bibinfo{person}{Edward Yang}, \bibinfo{person}{Zachary DeVito},
  \bibinfo{person}{Zeming Lin}, \bibinfo{person}{Alban Desmaison},
  \bibinfo{person}{Luca Antiga}, {and} \bibinfo{person}{Adam Lerer}.}
  \bibinfo{year}{2017}\natexlab{}.
\newblock \showarticletitle{Automatic differentiation in pytorch}.
\newblock  (\bibinfo{year}{2017}).
\newblock


\bibitem[Qin et~al\mbox{.}(2021)]%
        {qin2021world}
\bibfield{author}{\bibinfo{person}{Yuqi Qin}, \bibinfo{person}{Pengfei Wang},
  {and} \bibinfo{person}{Chenliang Li}.} \bibinfo{year}{2021}\natexlab{}.
\newblock \showarticletitle{The world is binary: Contrastive learning for
  denoising next basket recommendation}. In
  \bibinfo{booktitle}{\emph{Proceedings of the 44th International ACM SIGIR
  Conference on Research and Development in Information Retrieval}}.
  \bibinfo{pages}{859--868}.
\newblock


\bibitem[Qiu et~al\mbox{.}(2019)]%
        {qiu2019rethinking}
\bibfield{author}{\bibinfo{person}{Ruihong Qiu}, \bibinfo{person}{Jingjing Li},
  \bibinfo{person}{Zi Huang}, {and} \bibinfo{person}{Hongzhi Yin}.}
  \bibinfo{year}{2019}\natexlab{}.
\newblock \showarticletitle{Rethinking the item order in session-based
  recommendation with graph neural networks}. In
  \bibinfo{booktitle}{\emph{Proceedings of the 28th ACM international
  conference on information and knowledge management}}.
  \bibinfo{pages}{579--588}.
\newblock


\bibitem[Ren et~al\mbox{.}(2019)]%
        {ren2019repeatnet}
\bibfield{author}{\bibinfo{person}{Pengjie Ren}, \bibinfo{person}{Zhumin Chen},
  \bibinfo{person}{Jing Li}, \bibinfo{person}{Zhaochun Ren},
  \bibinfo{person}{Jun Ma}, {and} \bibinfo{person}{Maarten De~Rijke}.}
  \bibinfo{year}{2019}\natexlab{}.
\newblock \showarticletitle{Repeatnet: A repeat aware neural recommendation
  machine for session-based recommendation}. In
  \bibinfo{booktitle}{\emph{Proceedings of the AAAI Conference on Artificial
  Intelligence}}, Vol.~\bibinfo{volume}{33}. \bibinfo{pages}{4806--4813}.
\newblock


\bibitem[Rendle et~al\mbox{.}(2010)]%
        {2010Factorizing}
\bibfield{author}{\bibinfo{person}{Steffen Rendle}, \bibinfo{person}{Christoph
  Freudenthaler}, {and} \bibinfo{person}{Lars Schmidt-Thieme}.}
  \bibinfo{year}{2010}\natexlab{}.
\newblock \showarticletitle{Factorizing personalized Markov chains for
  next-basket recommendation}. In \bibinfo{booktitle}{\emph{Proceedings of the
  19th International Conference on World Wide Web, WWW 2010, Raleigh, North
  Carolina, USA, April 26-30, 2010}}.
\newblock


\bibitem[Sanh et~al\mbox{.}(2019)]%
        {sanh2019distilbert}
\bibfield{author}{\bibinfo{person}{Victor Sanh}, \bibinfo{person}{Lysandre
  Debut}, \bibinfo{person}{Julien Chaumond}, {and} \bibinfo{person}{Thomas
  Wolf}.} \bibinfo{year}{2019}\natexlab{}.
\newblock \showarticletitle{DistilBERT, a distilled version of BERT: smaller,
  faster, cheaper and lighter}.
\newblock \bibinfo{journal}{\emph{arXiv preprint arXiv:1910.01108}}
  (\bibinfo{year}{2019}).
\newblock


\bibitem[Shaw et~al\mbox{.}(2018)]%
        {Shaw2018Self}
\bibfield{author}{\bibinfo{person}{Peter Shaw}, \bibinfo{person}{Jakob
  Uszkoreit}, {and} \bibinfo{person}{Ashish Vaswani}.}
  \bibinfo{year}{2018}\natexlab{}.
\newblock \showarticletitle{Self-Attention with Relative Position
  Representations}. In \bibinfo{booktitle}{\emph{Proceedings of NAACL-HLT}}.
  \bibinfo{pages}{464--468}.
\newblock


\bibitem[Sun et~al\mbox{.}(2019)]%
        {sun2019bert4rec}
\bibfield{author}{\bibinfo{person}{Fei Sun}, \bibinfo{person}{Jun Liu},
  \bibinfo{person}{Jian Wu}, \bibinfo{person}{Changhua Pei},
  \bibinfo{person}{Xiao Lin}, \bibinfo{person}{Wenwu Ou}, {and}
  \bibinfo{person}{Peng Jiang}.} \bibinfo{year}{2019}\natexlab{}.
\newblock \showarticletitle{BERT4Rec: Sequential recommendation with
  bidirectional encoder representations from transformer}. In
  \bibinfo{booktitle}{\emph{Proceedings of the 28th ACM international
  conference on information and knowledge management}}.
  \bibinfo{pages}{1441--1450}.
\newblock


\bibitem[Tang and Wang(2018)]%
        {tang2018personalized}
\bibfield{author}{\bibinfo{person}{Jiaxi Tang} {and} \bibinfo{person}{Ke
  Wang}.} \bibinfo{year}{2018}\natexlab{}.
\newblock \showarticletitle{Personalized top-n sequential recommendation via
  convolutional sequence embedding}. In \bibinfo{booktitle}{\emph{Proceedings
  of the Eleventh ACM International Conference on Web Search and Data Mining}}.
  \bibinfo{pages}{565--573}.
\newblock


\bibitem[Van Den~Oord et~al\mbox{.}(2013)]%
        {van2013music}
\bibfield{author}{\bibinfo{person}{A{\"a}ron Van Den~Oord},
  \bibinfo{person}{Sander Dieleman}, {and} \bibinfo{person}{Benjamin
  Schrauwen}.} \bibinfo{year}{2013}\natexlab{}.
\newblock \showarticletitle{Deep content-based music recommendation}. In
  \bibinfo{booktitle}{\emph{Neural Information Processing Systems Conference
  (NIPS 2013)}}, Vol.~\bibinfo{volume}{26}. Neural Information Processing
  Systems Foundation (NIPS).
\newblock


\bibitem[Vaswani et~al\mbox{.}(2017)]%
        {vaswani2017attention}
\bibfield{author}{\bibinfo{person}{Ashish Vaswani}, \bibinfo{person}{Noam
  Shazeer}, \bibinfo{person}{Niki Parmar}, \bibinfo{person}{Jakob Uszkoreit},
  \bibinfo{person}{Llion Jones}, \bibinfo{person}{Aidan~N. Gomez},
  \bibinfo{person}{Lukasz Kaiser}, {and} \bibinfo{person}{Illia Polosukhin}.}
  \bibinfo{year}{2017}\natexlab{}.
\newblock \showarticletitle{Attention is All you Need}. In
  \bibinfo{booktitle}{\emph{Advances in NeurIPS 2017}}.
  \bibinfo{pages}{5998--6008}.
\newblock


\bibitem[Wang et~al\mbox{.}(2018)]%
        {wang2018billion}
\bibfield{author}{\bibinfo{person}{Jizhe Wang}, \bibinfo{person}{Pipei Huang},
  \bibinfo{person}{Huan Zhao}, \bibinfo{person}{Zhibo Zhang},
  \bibinfo{person}{Binqiang Zhao}, {and} \bibinfo{person}{Dik~Lun Lee}.}
  \bibinfo{year}{2018}\natexlab{}.
\newblock \showarticletitle{Billion-scale commodity embedding for e-commerce
  recommendation in alibaba}. In \bibinfo{booktitle}{\emph{Proceedings of the
  24th ACM SIGKDD International Conference on Knowledge Discovery \& Data
  Mining}}. \bibinfo{pages}{839--848}.
\newblock


\bibitem[Wu et~al\mbox{.}(2019)]%
        {wu2019session}
\bibfield{author}{\bibinfo{person}{Shu Wu}, \bibinfo{person}{Yuyuan Tang},
  \bibinfo{person}{Yanqiao Zhu}, \bibinfo{person}{Liang Wang},
  \bibinfo{person}{Xing Xie}, {and} \bibinfo{person}{Tieniu Tan}.}
  \bibinfo{year}{2019}\natexlab{}.
\newblock \showarticletitle{Session-based recommendation with graph neural
  networks}. In \bibinfo{booktitle}{\emph{Proceedings of the AAAI conference on
  artificial intelligence}}, Vol.~\bibinfo{volume}{33}.
  \bibinfo{pages}{346--353}.
\newblock


\bibitem[Xie et~al\mbox{.}(2020)]%
        {xie2020contrastive}
\bibfield{author}{\bibinfo{person}{Xu Xie}, \bibinfo{person}{Fei Sun},
  \bibinfo{person}{Zhaoyang Liu}, \bibinfo{person}{Shiwen Wu},
  \bibinfo{person}{Jinyang Gao}, \bibinfo{person}{Bolin Ding}, {and}
  \bibinfo{person}{Bin Cui}.} \bibinfo{year}{2020}\natexlab{}.
\newblock \showarticletitle{Contrastive learning for sequential
  recommendation}.
\newblock \bibinfo{journal}{\emph{arXiv preprint arXiv:2010.14395}}
  (\bibinfo{year}{2020}).
\newblock


\bibitem[Xu et~al\mbox{.}(2019)]%
        {xu2019graph}
\bibfield{author}{\bibinfo{person}{Chengfeng Xu}, \bibinfo{person}{Pengpeng
  Zhao}, \bibinfo{person}{Yanchi Liu}, \bibinfo{person}{Victor~S Sheng},
  \bibinfo{person}{Jiajie Xu}, \bibinfo{person}{Fuzhen Zhuang},
  \bibinfo{person}{Junhua Fang}, {and} \bibinfo{person}{Xiaofang Zhou}.}
  \bibinfo{year}{2019}\natexlab{}.
\newblock \showarticletitle{Graph Contextualized Self-Attention Network for
  Session-based Recommendation.}. In \bibinfo{booktitle}{\emph{IJCAI}},
  Vol.~\bibinfo{volume}{19}. \bibinfo{pages}{3940--3946}.
\newblock


\bibitem[Yuan et~al\mbox{.}(2020)]%
        {yuan2020future}
\bibfield{author}{\bibinfo{person}{Fajie Yuan}, \bibinfo{person}{Xiangnan He},
  \bibinfo{person}{Haochuan Jiang}, \bibinfo{person}{Guibing Guo},
  \bibinfo{person}{Jian Xiong}, \bibinfo{person}{Zhezhao Xu}, {and}
  \bibinfo{person}{Yilin Xiong}.} \bibinfo{year}{2020}\natexlab{}.
\newblock \showarticletitle{Future data helps training: Modeling future
  contexts for session-based recommendation}. In
  \bibinfo{booktitle}{\emph{Proceedings of The Web Conference 2020}}.
  \bibinfo{pages}{303--313}.
\newblock


\bibitem[Zhou et~al\mbox{.}(2019)]%
        {dien}
\bibfield{author}{\bibinfo{person}{Guorui Zhou}, \bibinfo{person}{Na Mou},
  \bibinfo{person}{Ying Fan}, \bibinfo{person}{Qi Pi}, \bibinfo{person}{Weijie
  Bian}, \bibinfo{person}{Chang Zhou}, \bibinfo{person}{Xiaoqiang Zhu}, {and}
  \bibinfo{person}{Kun Gai}.} \bibinfo{year}{2019}\natexlab{}.
\newblock \showarticletitle{Deep interest evolution network for click-through
  rate prediction}. In \bibinfo{booktitle}{\emph{Proceedings of the AAAI
  conference on artificial intelligence}}, Vol.~\bibinfo{volume}{33}.
  \bibinfo{pages}{5941--5948}.
\newblock


\bibitem[Zhou et~al\mbox{.}(2020)]%
        {zhou2020s3}
\bibfield{author}{\bibinfo{person}{Kun Zhou}, \bibinfo{person}{Hui Wang},
  \bibinfo{person}{Wayne~Xin Zhao}, \bibinfo{person}{Yutao Zhu},
  \bibinfo{person}{Sirui Wang}, \bibinfo{person}{Fuzheng Zhang},
  \bibinfo{person}{Zhongyuan Wang}, {and} \bibinfo{person}{Ji{-}Rong Wen}.}
  \bibinfo{year}{2020}\natexlab{}.
\newblock \showarticletitle{S3-Rec: Self-Supervised Learning for Sequential
  Recommendation with Mutual Information Maximization}. In
  \bibinfo{booktitle}{\emph{Proceedings of CIKM 2020}}.
  \bibinfo{pages}{1893--1902}.
\newblock


\bibitem[Zhou et~al\mbox{.}(2022)]%
        {zhou2022filter}
\bibfield{author}{\bibinfo{person}{Kun Zhou}, \bibinfo{person}{Hui Yu},
  \bibinfo{person}{Wayne~Xin Zhao}, {and} \bibinfo{person}{Ji-Rong Wen}.}
  \bibinfo{year}{2022}\natexlab{}.
\newblock \showarticletitle{Filter-enhanced MLP is All You Need for Sequential
  Recommendation}.
\newblock \bibinfo{journal}{\emph{arXiv preprint arXiv:2202.13556}}
  (\bibinfo{year}{2022}).
\newblock


\bibitem[Zhu et~al\mbox{.}(2020)]%
        {zhu2020modeling}
\bibfield{author}{\bibinfo{person}{Jing Zhu}, \bibinfo{person}{Yanan Xu}, {and}
  \bibinfo{person}{Yanmin Zhu}.} \bibinfo{year}{2020}\natexlab{}.
\newblock \showarticletitle{Modeling Long-Term and Short-Term Interests with
  Parallel Attentions for Session-Based Recommendation}. In
  \bibinfo{booktitle}{\emph{International Conference on Database Systems for
  Advanced Applications}}. Springer, \bibinfo{pages}{654--669}.
\newblock


\end{thebibliography}
